\documentclass[11pt,oneside,english]{amsart}
\usepackage{mathpazo}
\usepackage[scaled=0.92]{helvet}
\usepackage{courier}
\usepackage[T1]{fontenc}
\usepackage[a4paper]{geometry}
\usepackage{array}
\usepackage{float}
\usepackage{units}
\usepackage{multirow}
\usepackage{amstext}
\usepackage{amsthm}
\usepackage{amssymb}
\usepackage{graphicx}
\usepackage{setspace}
\usepackage[authoryear]{natbib}
\usepackage{babel}
\usepackage{color}
\usepackage[ruled]{algorithm2e}
\usepackage{wasysym}
\SetKwInput{KwInput}{Input}
\SetKwInput{KwOutput}{Output}
\usepackage{lineno}
\usepackage{hyperref}
\hypersetup{
    colorlinks = false
}

\makeatletter

\numberwithin{equation}{section}
\numberwithin{figure}{section}
\numberwithin{table}{section}

\newtheorem{theorem}{Theorem}[section]
\newtheorem{corollary}{Corollary}[section]
\theoremstyle{plain}

\newtheorem{lemma}{Lemma}[section]

\numberwithin{equation}{section}

\makeatother

\begin{document}

\title[Subsampling MCMC]{Subsampling MCMC - an introduction for the survey statistician}

\author{Matias Quiroz} 
\author{Mattias Villani} 
\author{Robert Kohn \\ Minh-Ngoc Tran}
\author{Khue-Dung Dang}
 
\thanks{Quiroz, Kohn and Dang: Australian School of Business, University of New South Wales. Villani: Department of Statistics, Stockholm University and Department of Computer and Information Science, Link\"oping University. Tran: Discipline of
Business Analytics, University of Sydney.}
\begin{abstract}
The rapid development of computing power and efficient Markov Chain Monte Carlo (MCMC) simulation algorithms have revolutionized Bayesian statistics, making it a highly practical inference method in applied work. However, MCMC algorithms tend to be computationally demanding, and are particularly slow for large datasets. Data subsampling has recently been suggested as a way to make MCMC methods scalable on massively large data, utilizing efficient sampling schemes and estimators from the survey sampling literature. These developments tend to be unknown by many survey statisticians who traditionally work with non-Bayesian methods, and rarely use MCMC. Our article explains the idea of data subsampling in MCMC by reviewing one strand of work, Subsampling MCMC, a so called pseudo-marginal MCMC approach to speeding up MCMC through data subsampling. The review is written for a survey statistician without previous knowledge of MCMC methods since our aim is to motivate survey sampling experts to contribute to the growing Subsampling MCMC literature. \\

\noindent \textit{AMS (2000) subject classification}. Primary 62-02; Secondary 62D05. 
\end{abstract}

\maketitle

\section{Introduction}
The key drivers behind the widespread adoption of Bayesian inference in the last three decades have been the rapid improvements in computing power and the availability of powerful user-friendly simulation algorithms. The family of Markov Chain Monte Carlo (MCMC) sampling methods \citep{brooks2011handbook}, and in particular the Metropolis-Hastings algorithm \citep{metropolis1953equation,hastings1970}, quickly became the method of choice for practitioners for simulating from complex posterior distributions. MCMC opened up the possibility of routine analysis of highly complex models with limited algorithmic tuning. MCMC sampling was also fast enough for most problems, and at first it seemed that the problem of computational intractability that had hindered early Bayesians had been solved once and for all.

Meanwhile, it became apparent that MCMC was too slow in certain specialized areas where particular problems still had practioners waiting for days or even weeks for MCMC to deliver the results. For example, MCMC is too slow for many high-dimensional spatial problems where the INLA approximations \citep{rue2009approximate} quickly gained popularity. Massive datasets in technology led to fast Variational Bayes (VB) approximations \citep{jordan1999introduction, blei2017variational} and Expectation Propagation (EP) \citep{minka2001expectation} in the machine learning field. The tension with MCMC for big data problems in the machine learning community is now present in many other scientific discplines in the natural and social sciences, and, with increasing text digitalization, also in the humanities. In the current big data era, MCMC is often too slow and is, as a result, increasingly being replaced by other approximate methods. This is unfortunate since, unlike other methods, MCMC samples are guaranteed to converge to the posterior distribution if the MCMC sampler performs adequately. Although there is exciting new work with flexible simulation based VB methods (see \citealp{blei2017variational} for a recent review), it is fair to say that VB is still less accurate than MCMC and does not come with practical error bounds. Moreover, it is often very time consuming to obtain good VB approximations for new complex models.

To deal with the challenges of massive datasets, there has been a recent push to develop scalable MCMC samplers. This work has followed two main paths: i) Distributed MCMC and ii) Subsampling MCMC. Distributed MCMC is inspired by the MapReduce scheme \citep{dean2008mapreduce} where the data is partitioned and distributed to different machines. MCMC is then run separately on each machine to obtain a subposterior for each partition in a parallel and distributed manner. The key question is then how to combine these subposteriors into a single posterior for all the data; see \cite{scott2016bayes}, \cite{neiswanger2013asymptotically}, \cite{minsker2014scalable}, \cite{wang2013parallel} and \cite{nemeth2018merging} for some attempts. Subsampling MCMC instead focuses on taking random subsamples of the data in each MCMC iteration. The FireFly Monte Carlo algorithm in \cite{maclaurin2014firefly} introduces an auxiliary variable for each observation which determines if it should be included
in the evaluation of the posterior; Gibbs sampling \citep{geman1984stochastic} is then used to switch between updates of the parameters and the auxiliary variables. \cite{korattikara2014austerity} and \cite{bardenet2014towards, bardenet2015markov} use increasingly larger subsets of the data until the accept-reject decision in MCMC can be taken with sufficiently high confidence. We refer to \cite{bardenet2015markov} for an excellent review of these and other subsampling approaches. After the publication of \cite{bardenet2015markov}, there has been interesting new progress on non-reversible MCMC for subsampling applications using continuous time piecewise deterministic Markov processes, see \cite{bierkens2016zig} and \cite{bouchard2018bouncy}. Moreover, a different approach using Noisy MCMC \citep{alquier2016noisy} and data subsampling is explored in \cite{maire2017informed}.

We will here focus on so called pseudo-marginal MCMC (PMCMC) methods where the likelihood evaluation is replaced by an unbiased estimate from a data subsample in each MCMC iteration \citep{andrieu2009pseudo}. Using a small subset to estimate the otherwise computationally costly likelihood in a big data setting can give dramatic speed-ups. As explained here, PMCMC has been shown to give samples from the correct posterior distribution even if the likelihood estimator is very noisy. However, as we demonstrate in this review, controlling the variability of the log of the likelihood estimator is absolutely crucial for the performance of Subsampling MCMC based on pseudo-marginal methods. This makes it important to introduce subsampling MCMC to survey sampling experts. The specific approach presented here has been developed in a series of papers \citep{quiroz2015speeding,quiroz2017SubsamplingDA,quiroz2016exact, dang2017hamiltonian} and should be of particular interest to survey statisticians since the estimation problem in our approach focuses on estimating the log-likelihood. The log-likelihood is usually a sum, and is therefore akin to a population total, the fundamental quantity in survey sampling. We also present a subsampling approach that directly estimates the likelihood unbiasedly \citep{quiroz2016exact}, which is usually a product; this is a less standard problem in survey sampling that may open up new challenges for survey statisticians. Finally, we note that estimating the log-likelihood based on data subsampling has also been explored in subsampling Sequential Monte Carlo (SMC) for static Bayesian models \citep{gunawan2018subsampling}. SMC \citep{doucet2001introduction} is a powerful alternative to MCMC which produces an estimate of the marginal likelihood, useful for model selection, as a byproduct. However, for brevity, this review focuses on MCMC.

The paper is organized as follows. The next section introduces the Metropolis-Hastings algorithm, and its extension to pseudo-marginal Metropolis-Hastings which can be used when the likelihood is replaced by an unbiased estimator. Section 3 gives details on estimators for the likelihood and their properties, and discusses several recently proposed variance reduction strategies such as using control variates and dependent subsamples. Section 3 also presents a promising approach for subsampling for Hamiltonian Monte Carlo (HMC) sampling which has recently been at the forefront in high-dimensional problems. The final section concludes. Appendix A summarizes the main algorithms and Appendix B gives some implementation details for our running illustrative example in the text.

\section{The Pseudo-Marginal Metropolis-Hastings (PMMH) Algorithm}

\subsection{The Metropolis-Hastings algorithm}\label{sub:MH}
Markov Chain Monte Carlo (MCMC) is a family of algorithms for random variate generation from potentially complicated multivariate distributions. MCMC simulates from a distribution $\pi(\boldsymbol{\theta})$, here taken as a Bayesian posterior distribution, by constructing a Markov Chain on the parameter space of $\boldsymbol{\theta}$ such that its invariant distribution is $\pi(\boldsymbol{\theta})$. Realizations from this Markov chain will therefore converge in distribution to $\pi(\boldsymbol{\theta})$ from any starting point of the Markov chain, such that after a burn-in period the path of the Markov chain is a dependent sample from $\pi(\boldsymbol{\theta})$. The celebrated \textit{Metropolis-Hastings (MH) algorithm} \citep{metropolis1953equation,hastings1970} in Algorithm \ref{alg:MH} in Appendix \ref{sec:algorithms}, is the most widely used MCMC algorithm. 

While the MH algorithm is valid for any proposal density $q(\boldsymbol{\theta}^\prime | \boldsymbol{\theta})$, where $\boldsymbol{\theta}$ is the current value of the parameter and $\boldsymbol{\theta}^\prime$ is its proposed value, the specific proposal used is crucial for the efficiency of the algorithm. The two most commonly used proposals are the Random Malk Metropolis (RWM) and the Independence sampler (IMH); see \cite{brooks2011handbook} for an introduction. The most common implementation of RWM uses a random walk proposal $q(\boldsymbol{\theta}^\prime | \boldsymbol{\theta})=N(\boldsymbol{\theta}, \kappa^2\Omega)$, where $\Omega$ captures the shape of posterior in an efficient implementation (often $\Omega$ is minus the inverse posterior Hessian or simply the identity matrix) and $\kappa$ is a tuning parameter. A small $\kappa$ is often needed to keep the acceptance probability reasonably large, and the algorithm therefore tends to traverse the parameter space very slowly. This is especially pronounced in high dimensions as the optimal $\kappa^2 = O(1/d)$, where $d$ is the number of parameters \citep{roberts1997weak}. The IMH sampler generates proposals independent of the current position: $q(\boldsymbol{\theta}^\prime | \boldsymbol{\theta}) = q(\boldsymbol{\theta}^\prime)$. Here it is crucial that $q(\boldsymbol{\theta}^\prime)$ is a fairly accurate approximation to the true posterior and that it has heavier tails, otherwise the sampler will generate long sequences of rejected draws, i.e. the sampler gets stuck for long spells. When the IMH proposal is a good approximation of the posterior, the sampler traverses the parameter space very swiftly.

\subsection{Estimating a computationally costly likelihood}\label{sec:costlyLikelihood}
The MH algorithm in Algorithm \ref{alg:MH} is extremely convenient for Bayesian computations since it does not require knowledge of the normalizing constant of the posterior, $p(\textbf{y})=\int p(\textbf{y} | \boldsymbol{\theta})p(\boldsymbol{\theta})d\boldsymbol{\theta}$, which is often intractable. Even so, there are many problems where the required evaluations of the likelihood $p(\textbf{y} | \boldsymbol{\theta})$ are also very costly, for example with large datasets or when the underlying probability model is a complex dynamical system, causing MH to be very slow. Moreover, for some models the likelihood can be intractable, e.g. in random effects models. Such situations are increasingly common in many of important applications and the slow execution of MH has prompted users to develop faster posterior approximation methods, for example variational Bayes \citep{blei2017variational} and expectation propagation \citep{gelman2017expectation}. While such methods are computationally attractive and steadily improving, they usually provide substantially less accurate approximations than MCMC.

A natural way to circumvent the problem of evaluating a costly likelihood $p(\textbf{y} | \boldsymbol{\theta})$ is to replace the likelihood by a computationally cheap estimate, $\hat p(\mathbf{y} | \boldsymbol{\theta})$. We will here illustrate this idea in two very different settings. 

\subsubsection*{Big data}
Consider first the big data case when we run the Metropolis-Hastings algorithm on a dataset with $n$ independent observations, with $n$ very large. Evaluating the likelihood is generally an $O(n)$ operation and can be very costly. A natural solution is to estimate the likelihood from a subsample of size $m$ obtained by simple random sampling. We first focus on estimating the \textit{log}-likelihood instead of the likelihood; the reason for estimating on the log-scale is that the log-likelihood is usually a sum and therefore equivalent to estimating a population total, a long studied problem in survey sampling \citep{sarndal2003model}. The log-likelihood for independent observations is
\begin{equation} \label{eq:sumLogLikelihood}
\ell (\mathbf{y} | \boldsymbol{\theta}) \equiv \log p(y_1,\ldots,y_n | \boldsymbol{\theta}) = \sum_{i=1}^n \ell _i(y_i | \boldsymbol{\theta}),
\end{equation} 
where $\ell_i(y_i | \boldsymbol{\theta}) = \log p(y_i | \boldsymbol{\theta} )$ is the \textit{log-likelihood contribution} of the $i$th observation. Let $u_1,\ldots,u_n$ be binary variables such that $u_i=1$ if observation $y_i$ is selected in the subsample, and zero otherwise. Assuming simple random sampling (SRS) without replacement, the usual unbiased estimator is of the simple form
\begin{equation} \label{eq:naiveEstimator}
    \hat\ell(\mathbf{y}|\boldsymbol{\theta}) \equiv \frac{n}{m} \sum_{i=1}^n \ell_i(y_i | \theta)u_i.
\end{equation}

While it is convenient from a survey sampling point of view to estimate the log-likelihood, we will see in Section \ref{sub:PMMH} that Subsampling MCMC actually requires an unbiased estimate of the \emph{likelihood} on the original scale. This entails estimating a product, which is a much less studied problem in survey sampling. In order to remain in the realm of survey sampling we can use the unbiased estimator for the log-likelihood in \eqref{eq:naiveEstimator} with a bias-correction to obtain an estimator for the likelihood of the form \citep{ceperley1999penalty,nicholls2012coupled}
\begin{equation}\label{eq:LikelihoodEstimator}
    \hat p(\mathbf{y} | \boldsymbol{\theta}) \equiv \exp \Big(\hat\ell(\mathbf{y}|\boldsymbol{\theta})- \sigma_{\hat \ell}^2(\boldsymbol{\theta})/2 \Big),
\end{equation}
where $\sigma_{\hat \ell}^2(\boldsymbol{\theta}) \equiv \mathrm{Var}(\hat\ell(\mathbf{y}|\boldsymbol{\theta}))$. This bias-correction is exact if i) $\hat\ell(\mathbf{y}|\boldsymbol{\theta})$ is normally distributed and ii) $\sigma_{\hat \ell}^2$ is known. In practice, $\sigma_{\hat \ell}^2$ is replaced by the usual sample estimate. We return to this issue in more detail in Section \ref{sub:approxApproach}.

Note that the log-likelihood can often be written as a sum even when the observations are not fully independent. The most straightforward example is longitudinal data where the time series of observations within a subject are typically dependent temporally, but the different subjects are independent. In this case the log-likelihood is a sum over subjects and we can estimate it from a subsample of subjects, rather than individual observations. Data with a direct Markovian structure can be handled similarly by subsampling an observation jointly with its relevant history, as is done in the block bootstrap for time series.

\subsubsection*{Random effects and importance sampling}
Another common setting where the likelihood is intractable, but can be estimated unbiasedly, are random effects models. As an example, consider a logistic regression with both fixed and random effects
$$ p(y_{it} | \mathbf{x}_{it}, \mathbf{w}_{it}, \beta, \alpha_i, \Sigma_{\alpha}) = \frac{\exp(\mathbf{x}_{it}^T\beta + \mathbf{w}_{it}^T\alpha_i)^{y_{it}}}{1+ \exp(\mathbf{x}_{it}^T\beta+ \mathbf{w}_{it}^T\alpha_i)},
$$ 
where $\mathbf{y}_i = (y_{i1},\ldots,y_{in_i})^T$ are $n_i$ observations for the $i$th subject, $\alpha_i \overset{iid}\sim  \mathrm{N}(0,\Sigma_{\alpha})$ are random effects of the covariates in $\mathbf{w}$, and $\mathbf{x}$ are covariates with fixed effects. 
The likelihood for a sample of $n$ observations with the random effects integrated out is then
\begin{equation}\label{eq:randomeffectslikelihood}
p(\mathbf{y}_1,\ldots,\mathbf{y}_n |  \mathbf{X}, \mathbf{W}, \beta, \Sigma_{\alpha}) = \prod_{i=1}^{n} \int_{\alpha_i}p(\mathbf{y}_i | \mathbf{X}_i, \mathbf{W}_i, \beta, \alpha_i) p(\alpha_i | \Sigma_{\alpha}) d \alpha_i,
\end{equation}
where
\begin{equation*}
    p(\mathbf{y}_i | \mathbf{X}_i, \mathbf{W}_i, \beta, \alpha_i, \Sigma_{\alpha}) = \prod_{t=1}^{n_i}p(y_{it} | \mathbf{x}_{it}, \mathbf{w}_{it}, \beta, \alpha_i).
\end{equation*}
The integrals in \eqref{eq:randomeffectslikelihood} are often intractable, but can be estimated unbiasedly by Monte Carlo integration, or importance sampling. Let $m_i$ denote the number of samples in the importance sampling estimate of each term, and $m=\sum_{i=1}^n m_i$ the total number of random numbers used to estimate the likelihood in \eqref{eq:randomeffectslikelihood}. Here importance sampling can be used to construct an unbiased estimate of the likelihood in random effects models. Similarly, for state space models, the particle filter gives an unbiased estimator of the likelihood using random particles, see \citet[Proposition 7.4.1]{del2004feynman} for the original result and \citet{pitt2012some} for an alternative proof.

It is important to highlight the randomness of the estimator so we write $\hat p(\mathbf{y} | \boldsymbol{\theta}, \mathbf{u})$, where $\mathbf{u} \sim p(\mathbf{u})$ are the random numbers used to form the estimate. In the large data setting, $\mathbf{u}$ is the vector of sample selection indicators discussed above and $p(\mathbf{u})$ is given by the simple random sampling design. More specifically, $\hat p(\mathbf{y} | \boldsymbol{\theta}, \mathbf{u})$ is given by  \eqref{eq:LikelihoodEstimator} with the log-likelihood estimate in \eqref{eq:naiveEstimator} showing the explicit dependence of the estimator on the random numbers $u_i$. In random effects models the $\mathbf{u}$ would instead be the random numbers used to approximate the intractable random effects integrals by Monte Carlo integration.

\subsection{The Pseudo-Marginal Metropolis-Hastings algorithm}\label{sub:PMMH}
\cite{andrieu2009pseudo}\\ prove the remarkable result that replacing the likelihood $p(\textbf{y} | \boldsymbol{\theta})$ in the MH algorithm with a noisy estimate $\hat p(\mathbf{y} | \boldsymbol{\theta}, \mathbf{u})$ still gives a sample from the posterior $\pi(\boldsymbol{\theta}) \propto p(\mathbf{y} | \boldsymbol{\theta})p(\boldsymbol{\theta})$ if the likelihood estimator $\hat p$ is positive and unbiased. This is done by defining an augmented target density that includes both $\boldsymbol{\theta}$ and $\mathbf{u}$ such that its marginal for $\boldsymbol{\theta}$ with $\mathbf{u}$ integrated out is the posterior of $\boldsymbol{\theta}$. The MH algorithm is run on this augmented target distribution and the $\mathbf{u}$ draws are not used for inference. It turns out that this so called \textit{pseudo-marginal} algorithm is exactly of the same form as the original MH algorithm, with the likelihood evaluation in the acceptance probability in each iteration replaced by its current estimate; see the Pseudo-Marginal Metropolis-Hastings (PMMH) in Algorithm \ref{alg:pseudoMargMH} for details. The idea of substituting the likelihood in MH with a noisy estimate appeared initially in physics \citep{lin2000noisy} and in genetics \citep{beaumont2003}.

Even though samples from PMMH with any unbiased positive likelihood estimator will converge to the posterior distribution, it turns out that having a low estimator variance is absolutely crucial for the efficiency of the standard PMMH sampler, see for example \cite{flury2011bayesian} and Section \ref{sub:controlVariatesCrucial}. An estimator with a large variance can easily lead to an accepted parameter draw with a large over-estimate of the likelihood; subsequent draws will be rejected until they also happen to be associated with another gross over-estimate. This causes the sampler to be stuck for long spells, making the MCMC algorithm very inefficient.

The variance of the likelihood estimate is controlled by $m$, the number of subsamples in the subsampling setting, or the number of draws in importance sampling estimators. An $m$ that is too small inflates the variance of the likelihood estimator and gives an inefficient sampler. An $m$ that is too large gives an unnecessarily precise estimator at an excessive computational cost. The optimal $m$ finds the right balance between MCMC efficiency and computational cost, and is usually derived under the assumption that the cost of a single MCMC iteration is proportional to $1/\mathrm{Var}(\log \hat p( \mathbf{y} | \boldsymbol{\theta}, \mathbf{u}))$, see e.g. \cite{pitt2012some} for details. This cost must be balanced against the efficiency of the MCMC (which can be shown to increase with $m$, as we will illustrate later). The usual measure of MCMC sampling inefficiency for a given parameter $\theta$ is given by the \textit{Integrated AutoCorrelation Time} (IACT)
\begin{equation}\label{eq:IACT}
\mathrm{IACT} = 1 + 2 \sum_{k=0}^\infty \rho_k,
\end{equation}
where $\rho_k$ is the $k$th autocorrelation of the MCMC chain for $\theta$. In practice, the IACT is estimated using the spectral density evaluated at zero, see for example \cite{coda2006}.
We define the Computational Time (CT) for producing a sample equivalent to an iid draw
from the posterior distribution as
\begin{equation}\label{eq:ComputationalTime}
\mathrm{CT}(\sigma_{\log \hat p}^2) \equiv \mathrm{IACT}(\sigma_{\log \hat p}^2) \times \text{Time for a single MH iteration} \propto \frac{\mathrm{IACT}(\sigma_{\log \hat p}^2)}{\sigma_{\log \hat p}^2},
\end{equation}
where $\sigma_{\log \hat p}^2 \equiv \mathrm{Var}(\log \hat p( \mathbf{y} | \boldsymbol{\theta}, \mathbf{u}))$. We note that the IACT in \eqref{eq:IACT} becomes a function of the variance of the log of the likelihood estimator when implementing pseudo-marginal MCMC. Here we follow \cite{pitt2012some} and \cite{doucet2015efficient} in assuming that the cost of a single iteration is proportional to $m$, which in turn is inversely proportional to $\sigma_{\log \hat p}^2$. Depending on the assumptions made, and the choice of proposal distribution for $\boldsymbol{\theta}$, the optimal subsample size $m$ which minimizes CT is obtained by targeting a $\sigma_{\log \hat p}^2$ between $1$ and $3.3$ \citep{pitt2012some, doucet2015efficient,sherlock2015efficiency}. It is also known that CT is relatively flat over the interval $\sigma_{\log \hat p}^2 \in [1,3.3]$, but increases sharply outside this interval, in particular when $\sigma_{\log \hat p}^2$ is too large. We will illustrate some properties of the CT later in the text.

The definition of CT in \eqref{eq:ComputationalTime} is the one traditionally used in pseudo-marginal MCMC. In some of the Subsampling MCMC methods the focus is on estimating the log-likelihood, which is subsequently converted into an estimator of the likelihood by bias-correction, see \eqref{eq:LikelihoodEstimator}. The relevant Computational Time is then
\begin{equation}\label{eq:ComputationalTimeLogL}
\mathrm{CT}(\sigma_{\hat \ell}^2) \equiv \frac{\mathrm{IACT}(\sigma_{\hat \ell}^2)}{\sigma_{\hat \ell}^2},
\end{equation}
where $\sigma_{\hat \ell}^2 \equiv  \mathrm{Var}(\hat \ell( \mathbf{y} | \boldsymbol{\theta}, \mathbf{u}))$. The two definitions of CT are identical if $\sigma_{\hat \ell}^2(\boldsymbol{\theta})$ in \eqref{eq:LikelihoodEstimator} is known, and typically differ very little when $\sigma_{\hat \ell}^2(\boldsymbol{\theta})$ in \eqref{eq:LikelihoodEstimator} is replaced by a sample estimate $\hat \sigma_{\hat \ell}^2(\boldsymbol{\theta})$. To keep things simple, we will therefore use the same rule to set the subsample size to target $\sigma_{\hat \ell}^2 \approx 1$ when using subsampling based on estimating the log-likelihood.

\section{Subsampling for likelihood estimation}
The previous section described how an estimated likelihood can be used in a pseudo-marginal algorithm to sample from a posterior distribution. As long as the estimator is unbiased and nonnegative, and some non-onerous regularity conditions apply, the samples will converge in distribution to the target posterior based on the true likelihood function. This section discusses the importance of variance reduction and proposes alternative estimators from the survey literature and adapts them to the Subsampling MCMC context.

\begin{figure}
    \centering
    \includegraphics[height=7.5cm, angle = -90]{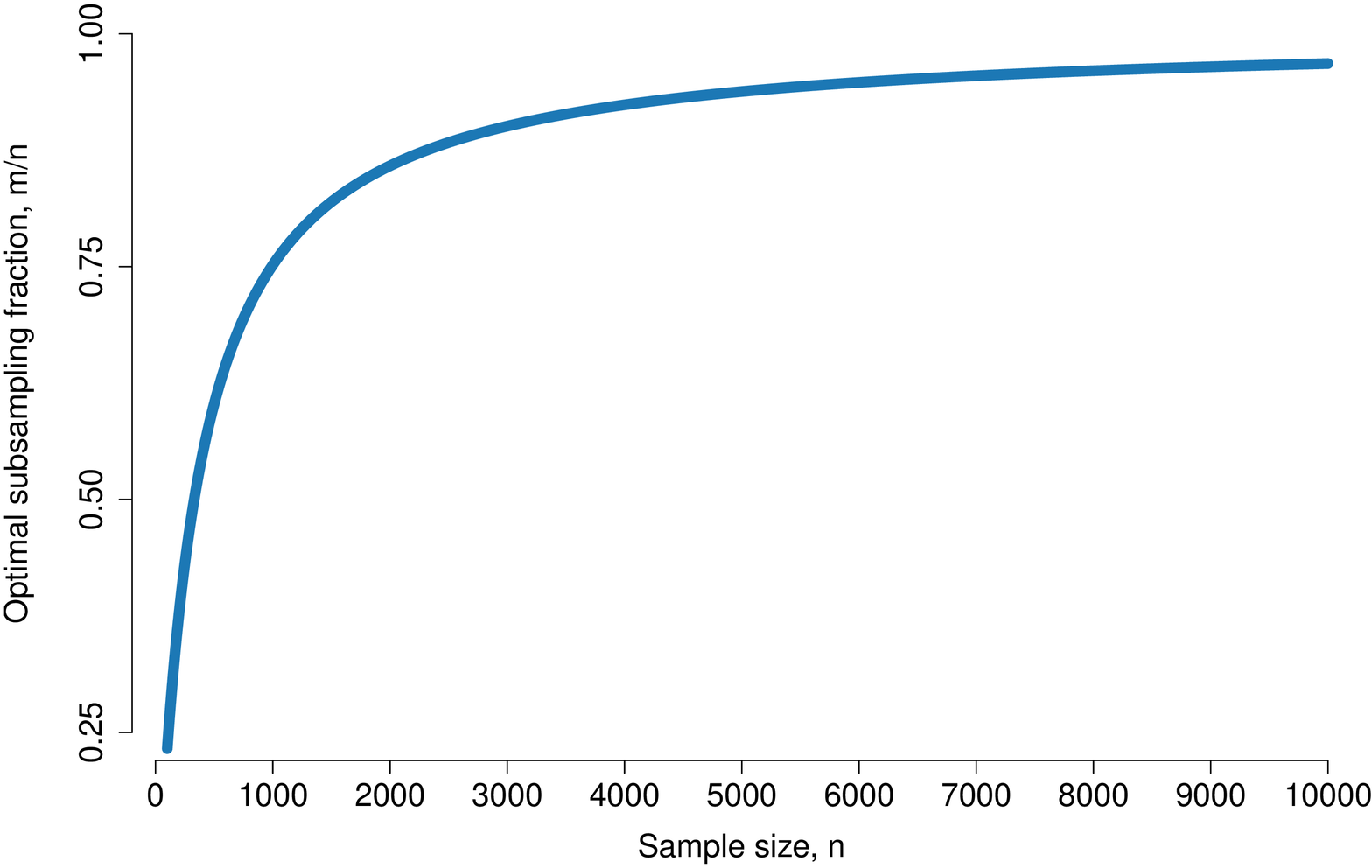}
    \includegraphics[height=7.5cm, angle = -90]{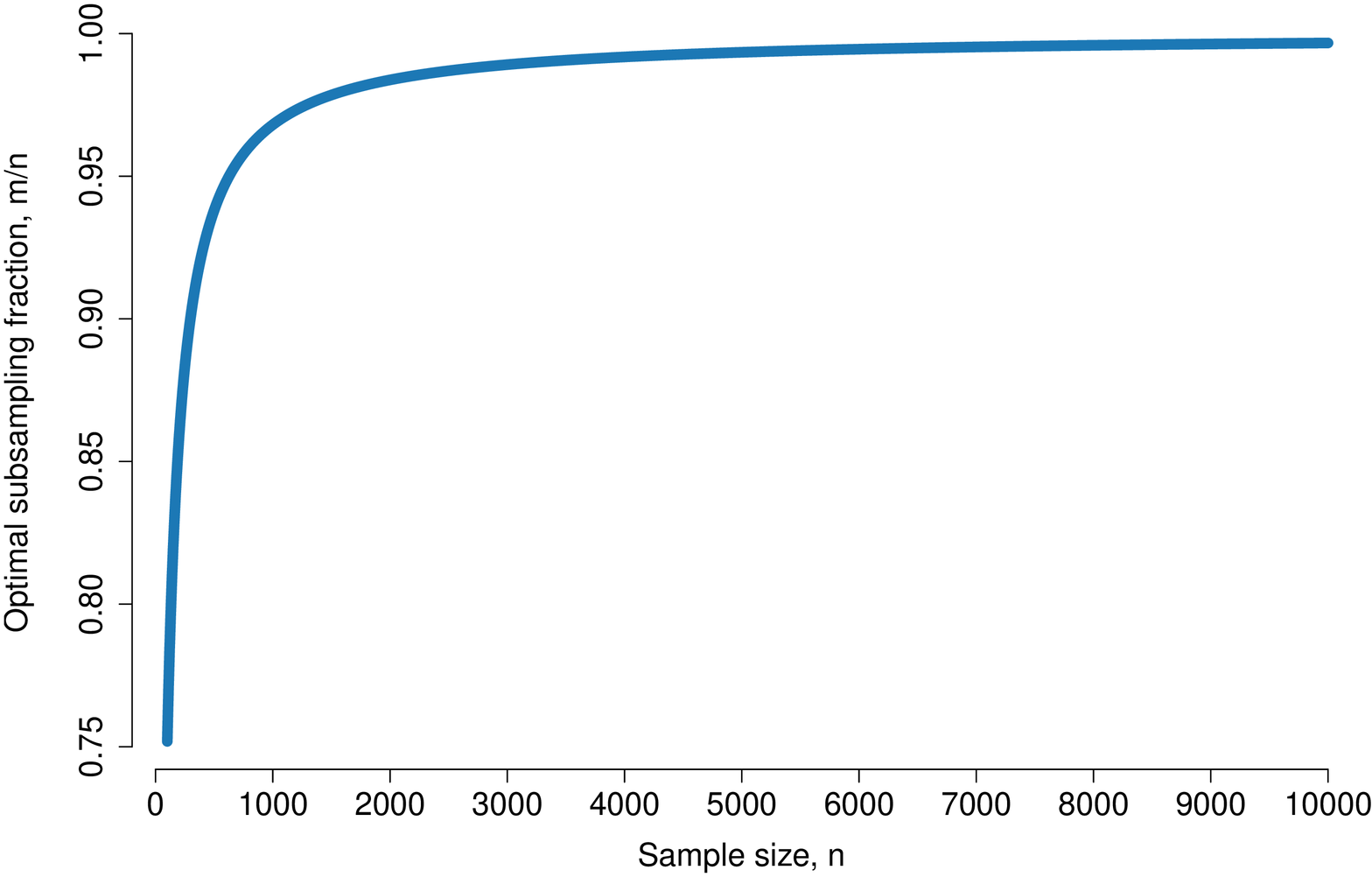}
    \caption{Optimal subsampling fractions ($m/n$) for SRS without replacement for $\sigma_{\ell_i}^2=1/100$ (left) and $\sigma_{\ell_i}^2=1/10$ (right), where $\sigma_{\ell_i}^2$ is the population variance in \eqref{eq:optimalm}.  The optimal subsample size ($m$) is set to target $\sigma_{\hat \ell}^2=3.3$.}
    \label{fig:samplingFractions}
\end{figure}

\subsection{Simple Random Sampling is by itself not useful for Subsampling MCMC}
We have already discussed that the optimal subsample size $m$ should target a variance of the log-likelihood estimator in the interval $\sigma_{\hat \ell}^2 \in [1,3.3]$. It turns out, however, that it is almost impossible in the subsampling setting to achieve a $\sigma_{\hat \ell}^2$ in that interval with Simple Random Sampling (SRS) without ending up with a sampling fraction $m/n$ very close to unity. To see this, note that the variance of the estimator in \eqref{eq:naiveEstimator} under the SRS design without replacement is \citep{sarndal2003model}
\begin{equation*}
    \sigma_{\hat \ell}^2 = \frac{n^2}{m}\Big( 1-\frac{m}{n} \Big) \sigma_{\ell_i}^2,
\end{equation*}
where $\sigma_{\ell_i}^2 \equiv \mathrm{Var}(\ell_i) = n^{-1}\sum_{i=1}^n (\ell_i-\bar \ell)^2$ is the population variance. Now, in order to target a given variance $\sigma_{\hat \ell}^2$, the subsample size must be
\begin{equation}\label{eq:optimalm}
    m = \frac{n^2\sigma_{\ell_i}^2}{n\sigma_{\ell_i}^2 + \sigma_{\hat \ell}^2}.
\end{equation}
Figure \ref{fig:samplingFractions} illustrates the optimal sampling fraction as a function of $n$ for two different values of $\sigma_{\ell_i}$ when the target is $\sigma_{\hat \ell}^2=3.3$. Note that this is the largest value $\sigma_{\hat \ell}^2$ among the recommended ones in the literature to keep the sampling fraction conservatively low here. The sampling fraction nevertheless quickly approaches unity, showing that SRS with the population total estimator in \eqref{eq:naiveEstimator} is not useful for Subsampling MCMC. An even more dramatic way of illustrating this is to consider sampling with replacement. SRS with replacement gives $\sigma_{\hat \ell}^2 = n^2 \sigma_{\ell_i}^2/m $ and the optimal $m$ grows as $O(n^2)$, which is clearly unacceptable.

The variance of the estimator when sampling without replacement is lower by the factor $1-m/n$ compared to the with-replacement case. This is a negligible improvement whenever $m\ll n$, which is the situation of interest here since otherwise subsampling would not be worthwhile. Since sampling with replacement is simpler to implement, and the implied independence makes the theory much easier to develop, this has been the preferred sampling method in the Subsampling MCMC literature. We will therefore use sampling with replacement throughout the paper. The sampling indicators $\mathbf{u} = (u_1,\ldots,u_m)$ are now random observation indices such that $\mathrm{Pr}(u_k=i)=1/n$ for $i=1,\ldots,n$ and the estimator in \eqref{eq:naiveEstimator} becomes 
\begin{equation} \label{eq:naiveEstimatorWithoutReplacement}
    \hat\ell(\mathbf{y}|\boldsymbol{\theta}) \equiv \frac{n}{m} \sum_{k=1}^m \ell_{u_k}(y_k | \boldsymbol{\theta}).
\end{equation}

\subsection{Efficient and scalable Subsampling MCMC using control variates}

\subsection*{The difference estimator}
Part of the problem with SRS is that the log-likelihood contributions $\ell_i(y_i|\boldsymbol{\theta})$ can vary quite dramatically over the observations, hence inflating the variance of the estimator. There are at least three main ways to deal with the heterogeneity of population elements. 

The first approach is stratified sampling with a higher sampling inclusion probability in the strata with largest units. This would ensure that most or all of the large $\ell_i(y_i|\boldsymbol{\theta})$ enter the sample. However, it turns out that stratified sampling tends to produce a variance that is too large for efficient Subsampling MCMC.

The second approach, proposed for Subsampling MCMC in the first version of \cite{quiroz2015speeding} (see \citealp{quiroz1stversion} for the first version), is to use probability-proportional-to-size (PPS) sampling that assigns higher inclusion probabilities to larger units \citep{sarndal2003model}. To implement PPS (or $\pi PS$ in the case of sampling without replacement) we need to approximate the size of $\ell_i(y_i|\boldsymbol{\theta})$ for all observations. In order to gain in computational speed from subsampling, those size measures must clearly be cheaper to compute than the $\ell_i(y_i|\boldsymbol{\theta})$, and such size measures are proposed in \citep{quiroz1stversion}. Nevertheless, the computational complexity of the subsampling algorithm remains $O(n)$.

The third approach is proposed in \cite{quiroz2015speeding} and amounts to subtracting an approximation $q_i(\boldsymbol{\theta})$ from each $\ell_i(y_i|\boldsymbol{\theta})$ such for each $\boldsymbol{\theta}$ that the new population $d_i(\boldsymbol{\theta}) =  \ell_i(y_i | \boldsymbol{\theta}) - q_i(\boldsymbol{\theta}) $ is more homogeneous in size than $\ell_i(y_i|\boldsymbol{\theta})$. Formally, we use simple random sampling and the \emph{difference estimator} \citep{sarndal2003model}
\begin{equation}  \label{eq:diffEstimator}
    \hat\ell_{DE}(\mathbf{y}|\boldsymbol{\theta},\mathbf{u}) \equiv \sum_{i=1}^n q_i(\boldsymbol{\theta}) + \frac{n}{m} \sum_{k=1}^m d_{u_k}(\boldsymbol{\theta}),
\end{equation}
with $q_i(\boldsymbol{\theta})$ is a potentially crude approximation to $\ell_i(y_i | \boldsymbol{\theta})$ for $i=1,\ldots,n$. It is easy to show that  $\hat\ell_{DE}(\mathbf{y}|\boldsymbol{\theta})$ is unbiased for any $q(\boldsymbol{\theta})$. The approximation $q(\boldsymbol{\theta})$ plays the same normalizing role as control variates in importance sampling \citep{hammersley1964} and we will use this term here. 

\subsection*{Parameter-expanded control variates}
A natural way of constructing control variates is by a Taylor expansion of $\ell (\mathbf{y}_i|\boldsymbol{\theta})$, $i=1\ldots,n$, around some central value $\boldsymbol{\theta} = \boldsymbol{\theta}^\star$ \citep{bardenet2015markov}
\begin{equation} \label{eq:paramControlVariates}
    \ell(\mathbf{y}_i|\boldsymbol{\theta}) \approx \ell (\mathbf{y}_i | \boldsymbol{\theta}^\star) 
    +  (\boldsymbol{\theta} - \boldsymbol{\theta}^\star)^T\nabla_{\boldsymbol{\theta}} \ell(\mathbf{y}_i|\boldsymbol{\theta})_{| \boldsymbol{\theta}=\boldsymbol{\theta}^\star}
    +  \frac{1}{2} (\boldsymbol{\theta} - \boldsymbol{\theta}^\star )^T\nabla_{\boldsymbol{\theta}\boldsymbol{\theta}^T}^2 \ell(\mathbf{y}_i|\boldsymbol{\theta})_{| \boldsymbol{\theta}=\boldsymbol{\theta}^\star} (\boldsymbol{\theta} - \boldsymbol{\theta}^\star),
\end{equation}
where $\nabla_{\boldsymbol{\theta}} \ell(\mathbf{y}_i|\boldsymbol{\theta})_{| \boldsymbol{\theta}=\boldsymbol{\theta}^\star}$ and $\nabla_{\boldsymbol{\theta}\boldsymbol{\theta}^T}^2 \ell(\mathbf{y}_i|\boldsymbol{\theta})_{| \boldsymbol{\theta}=\boldsymbol{\theta}^\star}$ are the gradient and Hessian with respect to $\boldsymbol{\theta}$ evaluated at $\boldsymbol{\theta} = \boldsymbol{\theta}^\star$, respectively. As argued in \cite{bardenet2015markov}, these \textit{parameter-expanded} control variates work very well when the posterior is tightly concentrated; asymptotic posterior concentration is guaranteed by the Bernstein von Mises theorem \citep{van1998asymptotic} and will be practically relevant in big data problems with many observations, but not too many parameters, i.e. so called \textit{tall data}. As discussed later, it also has good scaling properties with respect to $n$.

A crucial property of parameter-expanded covariates is that the sum $\sum_{i=1}^n q_i(\boldsymbol{\theta})$ in the difference estimator in \eqref{eq:diffEstimator} can be reduced from an $O(n)$ operation to an $O(1)$ operation since both $\sum_{i=1}^n\nabla_{\boldsymbol{\theta}} \ell(\mathbf{y}_i|\boldsymbol{\theta})_{| \boldsymbol{\theta}=\boldsymbol{\theta}^\star}$ and $\sum_{i=1}^n \nabla_{\boldsymbol{\theta}\boldsymbol{\theta}^T}^2 \ell(\mathbf{y}_i|\boldsymbol{\theta})_{| \boldsymbol{\theta}=\boldsymbol{\theta}^\star}$ are evaluated at $\boldsymbol{\theta}^\star$, and can therefore be pre-computed before starting the MCMC iterations.

\subsection*{Data-expanded control variates}

Let $\mathbf{z}_i$ be a vector with all observed data for the $i$th item. For example, in a regression setting, $\mathbf{z}_i=(y_i,\mathbf{x}_i^{T})^{T}$ would contain both the response variable $y$ and the covariates $\mathbf{x}$. Further, let $\ell(\mathbf{z}_i|\boldsymbol{\theta})$ denote log-likelihood contribution for the $i$th observation. The idea with the data-expanded control variates proposed in \cite{quiroz2015speeding} is that the $\ell(\mathbf{z}_i|\boldsymbol{\theta})$ tend to vary slowly across data space, and $\ell(\mathbf{z}_i|\boldsymbol{\theta})$ can therefore be approximated by $\ell(\mathbf{z}_{c_i}|\boldsymbol{\theta})$, where $\mathbf{z}_{c_i}$ is the nearest centroid in a pre-clustering of the data. Similarly to the parameter-expanded control variates, we can improve on this by using a Taylor expansion of $\ell(\mathbf{z}_i|\boldsymbol{\theta})$, but this time in data space around the centroid $\mathbf{z}_{c_i}$. The \textit{data-expanded} control variates are of the form
\begin{equation} \label{eq:dataControlVariates}
    \ell(\mathbf{z}_i|\boldsymbol{\theta}) \approx \ell (\mathbf{z}_{c_i} | \boldsymbol{\theta}) 
    +  (\mathbf{z}_{i}-\mathbf{z}_{c_i})^T \nabla_{\mathbf{z}} \ell(\mathbf{z}|\boldsymbol{\theta})_{| \mathbf{z} = \mathbf{z}_{c_i}} 
    +  \frac{1}{2} (\mathbf{z}_{i}-\mathbf{z}_{c_i})^T\nabla_{\mathbf{z}\mathbf{z}^T}^2 \ell(\mathbf{z}|\boldsymbol{\theta})_{| \mathbf{z}=\mathbf{z}_{c_i}} (\mathbf{z}_{i}-\mathbf{z}_{c_i}),
\end{equation}
where $\nabla_{\mathbf{z}} \ell(\mathbf{z}|\boldsymbol{\theta})_{| \mathbf{z} = \mathbf{z}_{c_i}}$ and $\nabla_{\mathbf{z}\mathbf{z}^T}^2 \ell(\mathbf{z}|\boldsymbol{\theta})_{| \mathbf{z}=\mathbf{z}_{c_i}}$ are the gradient and Hessian with respect to $\mathbf{z}$, both evaluated at $\mathbf{z} = \mathbf{z}_{c_i}$.

\cite{quiroz2015speeding} show that the complexity of $\sum_{i=1}^n q_i(\boldsymbol{\theta})$ is $O(K)$ for data-expanded control variates, where $K$ is the number of clusters and typically $K\ll n$. Hence, data-expanded control variates also give scalable algorithms since the number of clusters tends to grow very slowly with $n$.

\begin{figure}
    \centering
    \includegraphics[height=15cm, angle = -90]{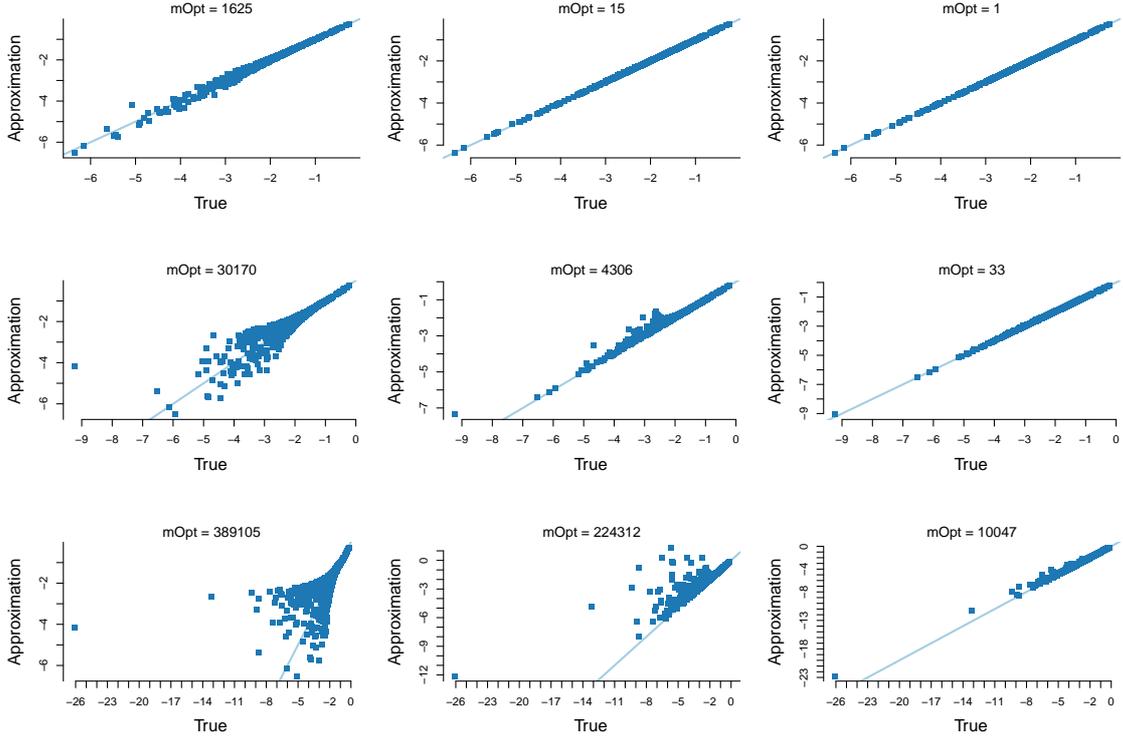}
    \caption{The accuracy of the parameter expanded control variates for the Poisson regression model in Eq. \eqref{eq:poisreg}. Each subgraph plots the true $\ell_i$ against the control variate for that observation. The three columns correspond to $0$, $1$ and $2$ terms in the Taylor expansion. The three rows correspond to different $\boldsymbol{\theta}$ that are increasingly distant from the Taylor expansion point $\boldsymbol{\theta}^\star$: i) $||\boldsymbol{\theta}-\boldsymbol{\theta}^\star||_2=0.025$ (top row), ii) $||\boldsymbol{\theta}-\boldsymbol{\theta}^\star||_2=0.1$ (middle row) and iii) $||\boldsymbol{\theta}-\boldsymbol{\theta}^\star||_2=0.25$ (bottom row), where $||\cdot||_2$ is the Euclidean norm. As a point of comparison, $\boldsymbol{\theta}$'s on the $50$\% posterior ellipsoid have values for $||\boldsymbol{\theta}-\boldsymbol{\theta}^\star||$ ranging between $0.013$ and $0.028$. The header of each subgraph displays the optimal subsample ($m_{opt}$) that gives the target variance $\mathrm{Var}(\hat \ell_{DE}) = 3.3$. The quality of the parameter expanded control variates depends on $\boldsymbol{\theta} - \boldsymbol{\theta}^\star$ being small.}
    \label{fig:paramexpandedAccuracy}
\end{figure}

\subsection*{Comparing control variates from parameter-expansion and data-expansion}
\begin{figure}
    \centering
    \includegraphics[height=15cm, angle = -90]{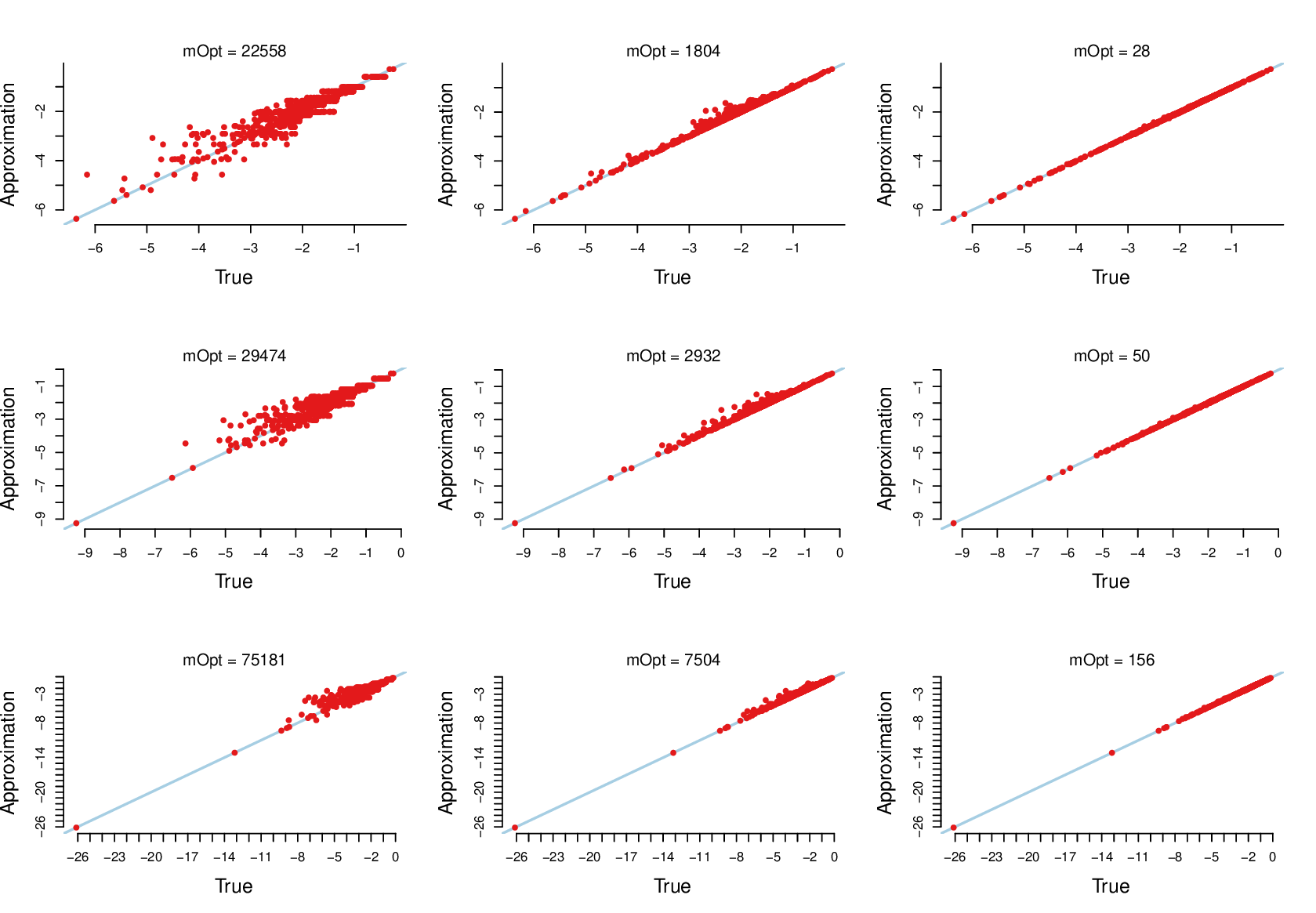}
    \caption{The accuracy of the data expanded control variates with 75 centroids for the Poisson regression model in Eq. \eqref{eq:poisreg}. Each subgraph plots the true $\ell_i$ against the control variate for that observation. The three columns correspond to $0$, $1$ and $2$ terms in the Taylor expansion. The three rows correspond to different $\boldsymbol{\theta}$ that are increasingly distant from the Taylor expansion point $\boldsymbol{\theta}^\star$: i) $||\boldsymbol{\theta}-\boldsymbol{\theta}^\star||_2=0.025$ (top row), ii) $||\boldsymbol{\theta}-\boldsymbol{\theta}^\star||_2=0.1$ (middle row) and iii) $||\boldsymbol{\theta}-\boldsymbol{\theta}^\star||_2=0.25$ (bottom row), where $||\cdot||_2$ is the Euclidean norm. As a point of comparison, $\boldsymbol{\theta}$'s on the $50$\% posterior ellipsoid have values for $||\boldsymbol{\theta}-\boldsymbol{\theta}^\star||$ ranging between $0.013$ and $0.028$. The header of each subgraph displays the optimal subsample  ($m_{opt}$) that gives the target variance $\mathrm{Var}(\hat \ell_{DE}) = 3.3$. The quality of the data expanded control variates is not sensitive to $\boldsymbol{\theta} - \boldsymbol{\theta}^\star$.}
    \label{fig:dataexpandedAccuracy}
\end{figure}

It is crucial to realize that our sampling problem is dynamic, in the sense that we will need estimates of $\hat \ell (\mathbf{y} | \boldsymbol{\theta})$ at every iteration of the pseudo-marginal MH algorithm, and $\boldsymbol{\theta}$ typically changes in every iteration. This means that we have sequence of survey sampling problems where the measurements on the population units, $\ell (\mathbf{y}_i | \boldsymbol{\theta}), i=1,\ldots,n$, change over time (MH iterations). Such situations also occur in real-world surveys \citep{steel2009design}, but Subsampling MCMC has not as yet used any of the methods proposed in the repeated surveys literature. We return to this dynamic survey sampling perspective when we discuss dependent subsampling in Section \ref{sec:CPM}. The fact that $\boldsymbol{\theta}$ changes over the iterations can cause problems for the parameter-expanded control variates, but does not significantly affect the data-expanded control variates. This is illustrated in Figures \ref{fig:paramexpandedAccuracy} and \ref{fig:dataexpandedAccuracy} where we plot the true $\ell (\mathbf{y}_i| \boldsymbol{\theta})$ against the two control variates for different number of terms in the Taylor expansions. For illustration purposes the underlying sample of $n=1000$ observations comes from a simple Poisson regression 
\begin{equation}\label{eq:poisreg}
    y_i | x_i \sim \mathrm{Pois}(\exp(\theta_0+\theta_1 x_i)),
\end{equation}
where $\boldsymbol{\theta} = (\theta_0,\theta_1)=(1,0.75)$, but the point we make holds generally. Figure \ref{fig:paramexpandedAccuracy} clearly shows that parameter-expansion around a static $\boldsymbol{\theta}^\star$ is problematic when the current $\boldsymbol{\theta}$ is far from $\boldsymbol{\theta}^\star$. Figure \ref{fig:dataexpandedAccuracy} shows that the data-expanded control variates remains relatively unaffected by movements in $\boldsymbol{\theta}$.

\begin{figure}
    \centering
    \includegraphics[height=7cm, angle = -90]{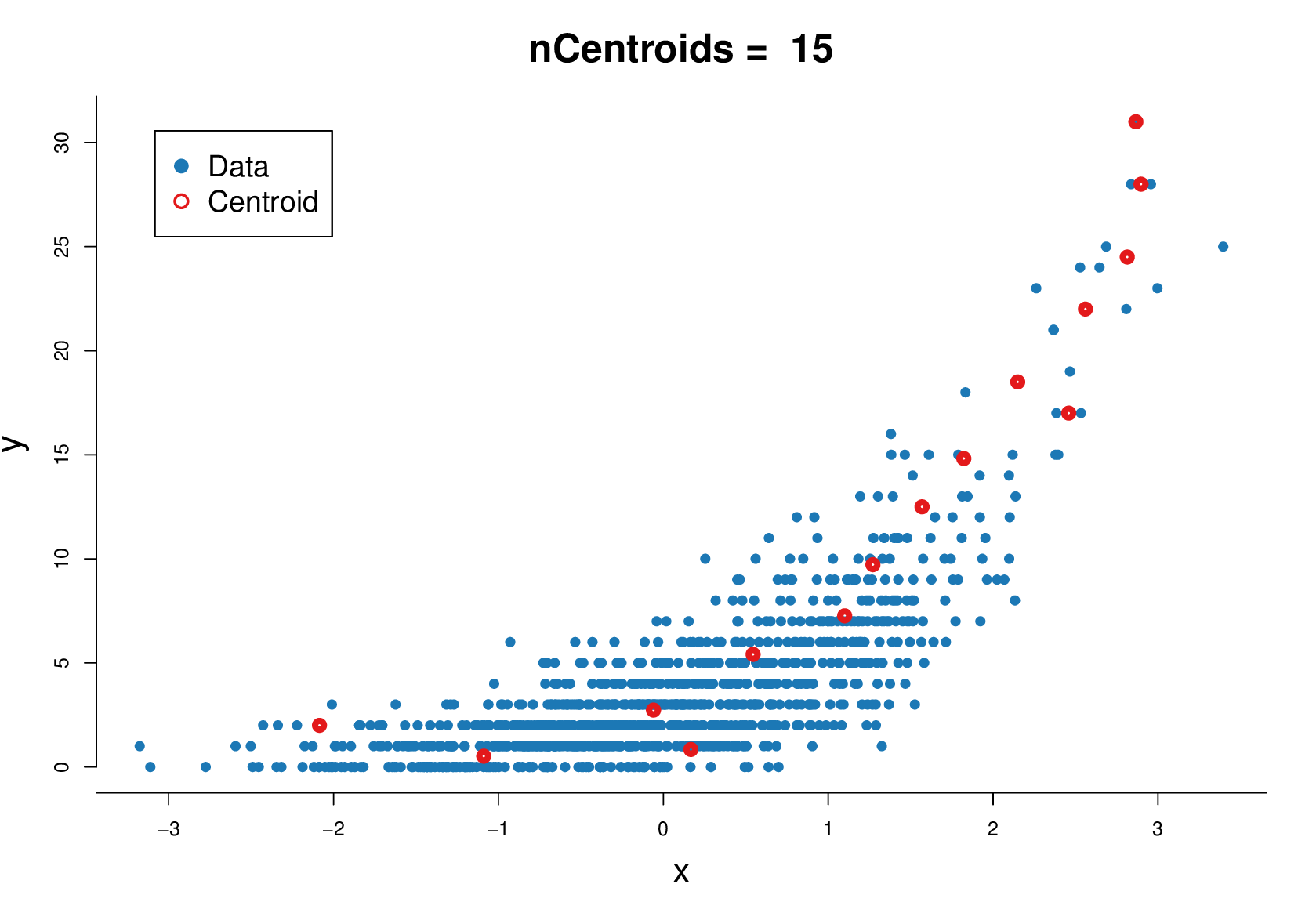}
    \includegraphics[height=7cm, angle = -90]{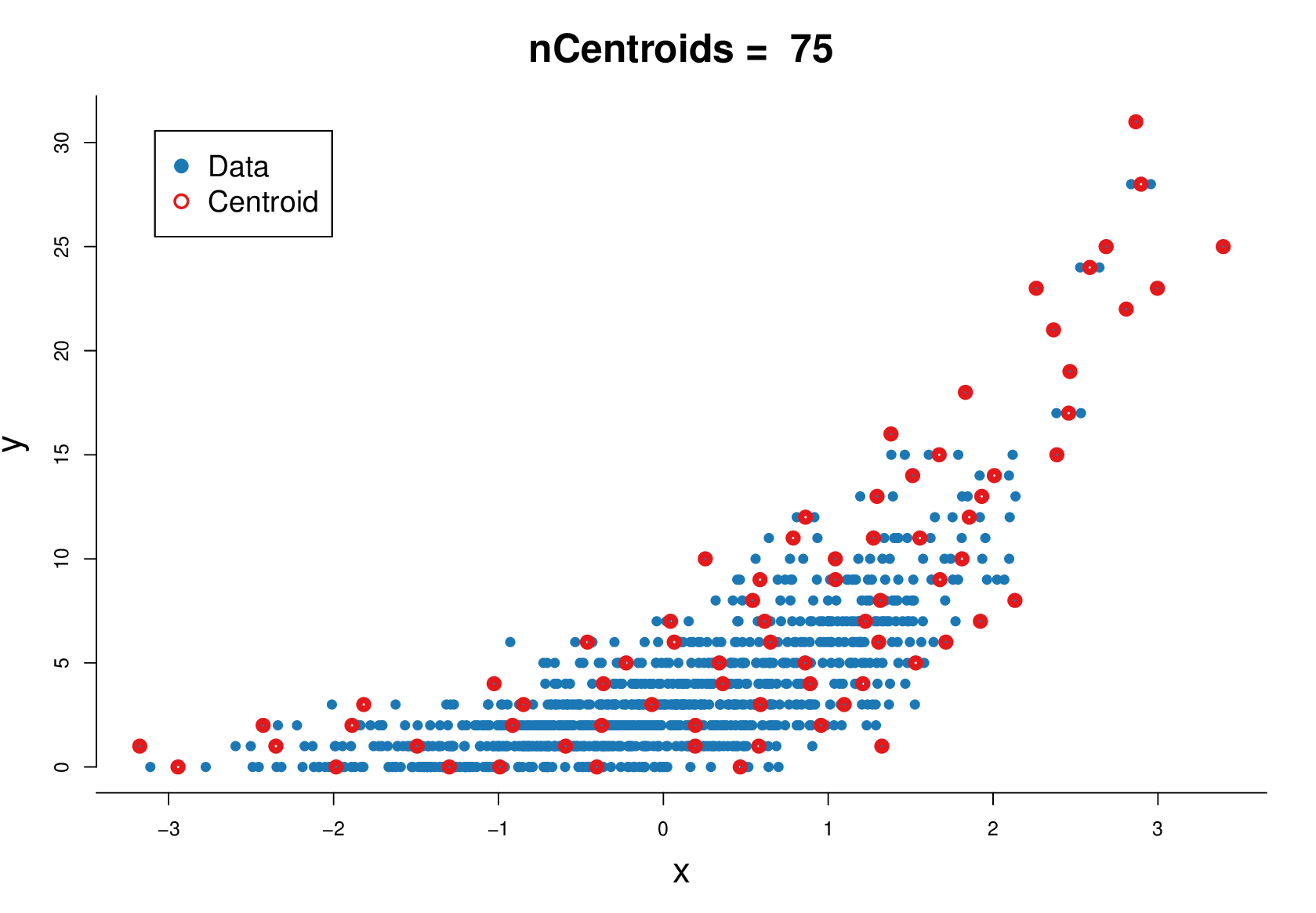}
    \caption{The data points in (x,y)-space and the centroids.}
    \label{fig:clustering}
\end{figure}

\begin{figure}
    \centering
    \includegraphics[height=15cm, angle = -90]{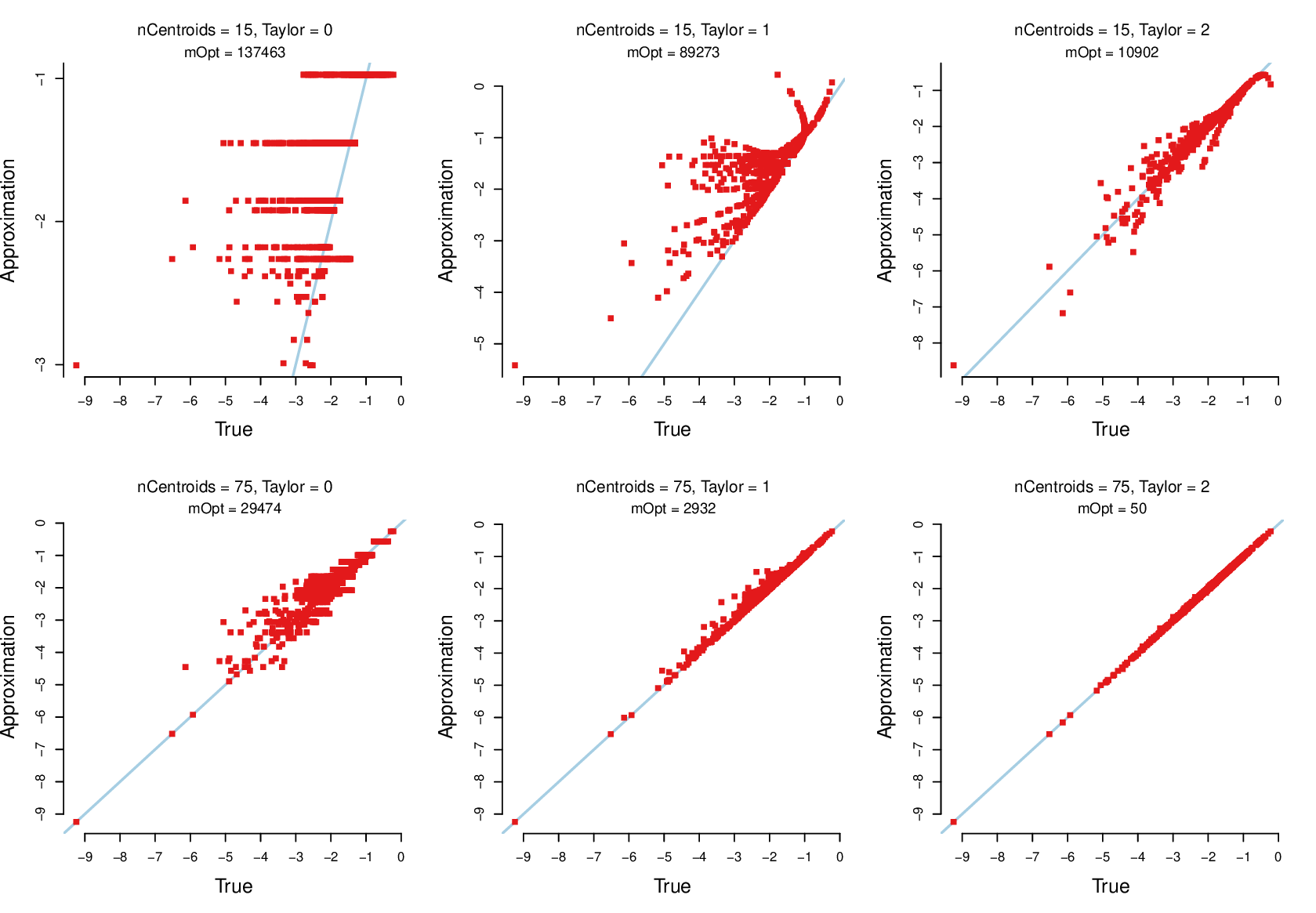}
    \caption{The accuracy of the data expanded control variates with different number of centroids for the Poisson regression model in Eq. \eqref{eq:poisreg} when $\boldsymbol{\theta}$ is such that $||\boldsymbol{\theta}-\boldsymbol{\theta}^\star||_2=0.1$, where $||\cdot||_2$ is the Euclidean distance. The three columns correspond to $0$, $1$ and $2$ terms in the Taylor expansion.  The header of each subgraph displays the optimal subsample  ($m_{opt}$) that gives the target variance $\mathrm{Var}(\hat \ell_{DE}) = 3.3$ and the order of the Taylor expansion (Taylor). The quality of the data expanded control variates deteriorates when the number of centroids is small, especially when using a lower order Taylor expansion.}
    \label{fig:dataexpandedCentroids}
\end{figure}

However, data-expanded control variates only give accurate approximations if enough centroids are used in the clustering; see Figures \ref{fig:clustering} and \ref{fig:dataexpandedCentroids} for an illustration. The curse of dimensionality makes this a limitation in higher dimensional data spaces since many observations will be quite far from their nearest centroid even when using a larger number of centroids. 

In summary, data-expanded control variates perform well for any $\boldsymbol{\theta}$, but do not scale well with the dimension of the data space. Parameter-expanded control variates scale well with dimension, but perform poorly when $\boldsymbol{\theta}$ is far from the expansion point $\boldsymbol{\theta}^\star$. \cite{quiroz2015speeding} therefore propose the strategy of starting the posterior sampling with data-expanded control variates and then switching over to parameter-expanded control variates when the sampler has reached a more central point in the posterior which can be used as $\boldsymbol{\theta}^\star$.

\subsection*{Asymptotic behavior with control variates}
We have shown that the optimal subsample size needs to grow as $O(n^2)$ when using simple random sampling with replacement in order to keep $\mathrm{Var}(\hat \ell (\mathbf{y}| \boldsymbol{\theta}))$ around unity; control variates can improve on this asymptotic rate. With control variates, the variance of the difference estimator in \eqref{eq:diffEstimator} is given by
\begin{equation}
    \mathrm{Var} \big( \hat\ell_{DE} \big) = \frac{n^2 \sigma^2_d(n)}{m},
\end{equation}
where $\sigma^2_d(n) \equiv (1/n)\sum_{i=1}^n (d_i-\bar d)^2$ is the variance of the finite population of differences. Note that we have made explicit that the accuracy of the control variates depends on $n$. As explained above, to obtain the optimal $m$ we need to ensure that $\mathrm{Var} \big( \hat\ell_{DE} \big)$ is $O(1)$, which requires understanding the behaviour of $\sigma^2_d(n)$ as $n \rightarrow \infty$. Lemma 2 in \cite{quiroz2015speeding} shows that 
\begin{equation}\label{eq:varDEest}
    \mathrm{Var} \big( \hat\ell_{DE} \big) = \frac{n^2 O(a_n^2)}{m},
\end{equation}
where
\begin{equation*}
    a_{n}(\boldsymbol{\theta}) \equiv 2\underset{i\in \{ 1,\ldots,n\}}{\max} | d_i(\boldsymbol{\theta})|.
\end{equation*}
The asymptotic behaviour of $a_{n}(\boldsymbol{\theta})$ depends on the type of control variate, and also on choices within a given control variate such as how the number of centroids grows with $n$ in the case of data-expanded control variates. We will focus here on the asymptotic properties of parameter-expanded control variates and refer to \cite{quiroz2015speeding} for results on the data-expanded case. 

Since the parameter-expanded control variate is based on a Taylor expansion around $\boldsymbol{\theta}^{\star}$, the rate at which its accuracy improves with $n$ is determined by the rate at which $||\boldsymbol{\theta} - \boldsymbol{\theta}^{\star}_n ||_2$ contracts, where we have made explicit that the expansion point $\boldsymbol{\theta}^{\star}_n$ typically depends on $n$. \cite{quiroz2015speeding} prove the following lemma.

\begin{lemma}\label{lemma:asymptotics4m}For the parameter-expanded control variates of second order we have
    $$a_n(\boldsymbol{\theta}) = ||\boldsymbol{\theta}-\boldsymbol{\theta}^\star_n||_2^3 \cdot O(1).$$
\end{lemma}
From the Bernstein-von Mises theorem \citep{chen1985asymptotic}, if $\boldsymbol{\theta}^\star_n$ is the posterior mode based on all data, we have $\sqrt{n}(\boldsymbol{\theta}-\boldsymbol{\theta}^\star_n) \overset{d}{\rightarrow} N(0,\tau^2)$ as $n\rightarrow\infty$. This implies that $\mathrm{Pr}(||\boldsymbol{\theta}-\boldsymbol{\theta}^\star_n||\leq K\tau/\sqrt n)$ will be close to unity for large enough $K$. We therefore have that $a_n(\boldsymbol{\theta})=O(n^{-3/2})$ for all $\boldsymbol{\theta} \in \{\boldsymbol{\theta}:||\boldsymbol{\theta}-\boldsymbol{\theta}^\star_n||\leq K\tau/\sqrt n\}$.
Hence, for such $\boldsymbol{\theta}$, the optimal subsample size that targets $\mathrm{Var} \big( \hat\ell_{DE} \big)=O(1)$ is, by \eqref{eq:varDEest}, $m=O(n^{-1})$, suggesting that Subsampling MCMC with parameter-expanded control variates scales extremely well to large datasets. There are at least three objections to this analysis, however. First, the conditions under which this optimality is derived requires that $m$ is large enough for $\hat \ell$ to be approximately normally distributed, so the optimal $m=O(n^{-1})$ is not attainable. Second, having control variates that expand around the posterior mode of $\boldsymbol{\theta}$ based on all data is not practical in large data settings. Third, as discussed in \cite{quiroz2015speeding}, setting $m=O(n^{-1})$ gives a PMMH algorithm that samples from a target distribution that deviates from the true posterior by an $O(n)$ factor, which is clearly not acceptable. A more practical approach with control variates based on the posterior mode from a small subset of the data is analyzed in \cite{quiroz2015speeding} and presented in Section \ref{sub:approxApproach} below.

\subsection*{Other control variates}\label{sub:otherControlVariates}
We have emphasized parameter- and data-expanded control variates as general and scalable solutions for variance reduction in Subsampling MCMC. However, many other control variates can be used in particular applications. For example, in many models the evaluation of the log-likelihood contributions $\ell(y_i | \boldsymbol{\theta})$ is very time-consuming because some aspect of the model needs to be solved numerically. The likelihood can then be costly also for smaller $n$. For example, an intractable integral may be approximated by Gaussian quadrature, a differential equation can be solved by the Runge-Kutta method, an optimum found by Newton's method. Any numerical method depends on tuning parameters which control the accuracy of the solution. A natural control variate can then be obtained from tuning parameters that give cruder, but much faster, evaluations of $\ell(y_i | \boldsymbol{\theta})$ (a coarse grid in numerical integration and in solving differential equations, a small number of Newton steps for optimization). The log-likelihood contributions for the sampled subset of observations are computed based on tuning parameters that give very accurate evaluations. Note however that for such control variates we need in general to evaluate the control variate for all $n$ observations (but $n$ may be small), so the algorithm will still run in $O(n)$ time, but with a much smaller cost for each MCMC iteration. 

\begin{figure}
    \centering
    \includegraphics[height=15cm, angle = -90]{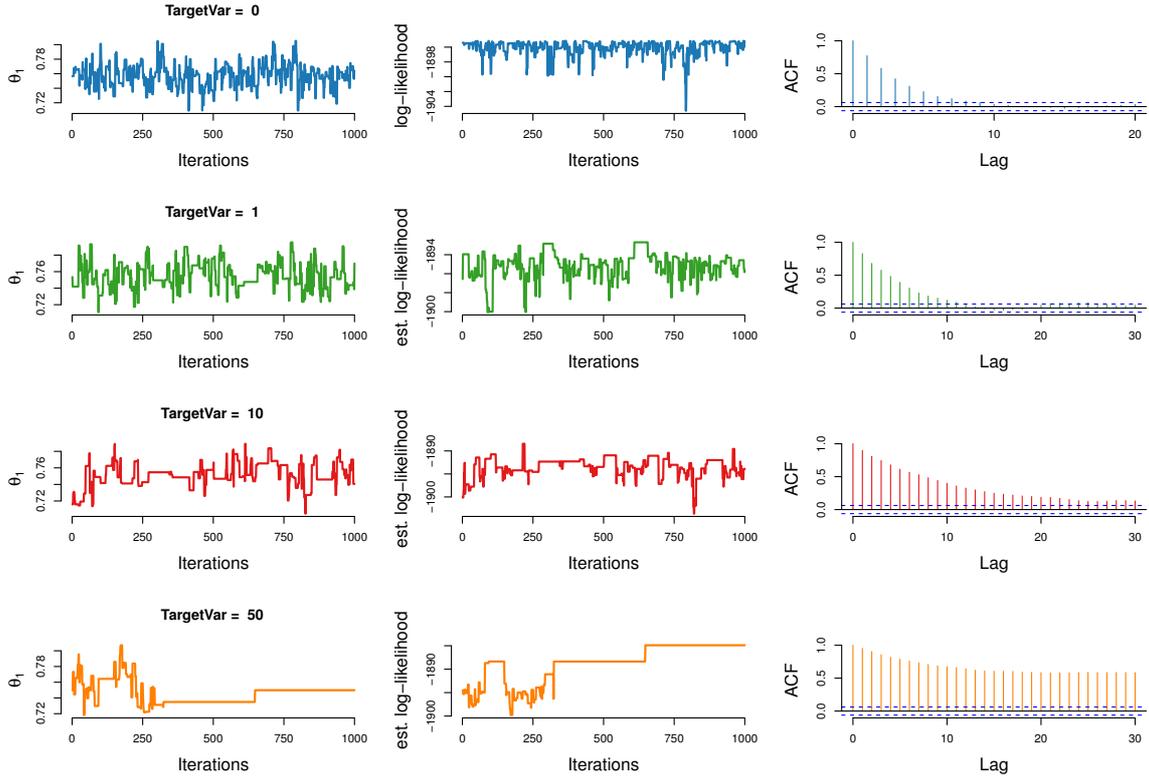}
    \caption{Sampling the posterior distribution of $\theta_1$ in the Poisson regression model in \eqref{eq:poisreg}. The figure shows, for different values of $\sigma_{\hat{\ell}}^2\in\{0,1,10,50\}$ on the four rows, the sampling chains (left), the estimates of the log-likelihood (middle) and estimates of the chain's autocorrelation $\rho_k$ (right).}
    \label{fig:subsamplingMCMC}
\end{figure}

\subsection{Control variates are crucial for the Integrated AutoCorrelation Time (IACT)}\label{sub:controlVariatesCrucial}
We have argued that control variates provide significant variance reduction for the log-likelihood estimator and that the MCMC sampling efficiency (as measured by the IACT) is a function of the variance. Figure \ref{fig:subsamplingMCMC} illustrates that when targeting a variance of one for $\hat \ell (\boldsymbol{\theta})$ (second row) our subsampling MCMC essentially behaves as the full data MCMC (first row). However, Subsampling MCMC with a large estimator variance ($\mathrm{Var}(\hat \ell (\boldsymbol{\theta}))=10$ in the third row and $\mathrm{Var}(\hat \ell (\boldsymbol{\theta}))=50$ in the fourth row) does not efficiently explore the posterior distribution of our Poisson regression example, and has a much greater tendency of getting stuck. This stickiness is also clearly borne out in the autocorrelation function of the MCMC draws in the right panel of Figure \ref{fig:subsamplingMCMC}.

\subsection{An approximate approach using bias-corrected log-likelihood estimators}\label{sub:approxApproach}

We have so far shown the importance of variance reduction of the log-likelihood estimator using control variates. The reason for focusing on estimators of the log-likelihood, rather than the likelihood, is that the log-likelihood is a sum, which is the usual aim in survey sampling, allowing us to exploit century-old experience in that area.

However, pseudo-marginal MCMC will only generate a sample from the correct posterior if the \textit{likelihood} is estimated by a positive unbiased estimator \citep{andrieu2009pseudo}. The difference estimator in \eqref{eq:diffEstimator} is unbiased for the log-likelihood, but biased for the likelihood. As discussed in Section \ref{sec:costlyLikelihood}, we can bias-correct the biased estimator $\exp(\hat\ell(\mathbf{y}|\boldsymbol{\theta}))$, where $\hat\ell(\mathbf{y}|\boldsymbol{\theta})$ is any unbiased estimator of the log-likelihood. In particular, using the difference estimator the bias-corrected estimator is of the form
\begin{equation}\label{eq:biasCorrectDEEstKnownSigma} 
     \exp \Big(\hat\ell_{DE}(\mathbf{y}|\boldsymbol{\theta})- \sigma_{\hat \ell_{DE}}^2(\boldsymbol{\theta})/2 \Big),
\end{equation}
where $\hat \ell_{DE}(\mathbf{y} | \boldsymbol{\theta})$ is the difference estimator in \eqref{eq:diffEstimator} and $\sigma_{\hat \ell_{DE}}^2=\mathrm{Var}(\hat\ell_{DE}(\mathbf{y}|\boldsymbol{\theta}))$. The estimator in \eqref{eq:biasCorrectDEEstKnownSigma} is unbiased if i) $\hat\ell_{DE}(\mathbf{y}|\boldsymbol{\theta})$ is normally distributed and ii) $\sigma_{\hat \ell_{DE}}^2$ is known. The assumption of normality can often be defended by a central limit theorem in the large $m$ setting (assuming also that $n$ grows). Even when $m$ is very small we have observed that $\hat\ell_{DE}(\mathbf{y}|\boldsymbol{\theta})$ is very close to normal since the control variates homogenize the population so that $d_i(\boldsymbol{\theta}) =  \ell(y_i | \boldsymbol{\theta}) - q_i(\boldsymbol{\theta}), i\in\{1,\ldots,n\}$, are usually distributed much more symmetrically and have lighter tails than the population of $\{\ell(y_i | \boldsymbol{\theta})\}_{i=1}^n$. Assuming that $\sigma_{\hat \ell_{DE}}^2$ is known is harder to defend since knowing $\sigma_{\hat \ell_{DE}}^2$ requires computing $d_i(\boldsymbol{\theta})$ for all $n$ observations. The approach in \cite{quiroz2015speeding} replaces $\sigma_{\hat \ell_{DE}}^2$ in \eqref{eq:biasCorrectDEEstKnownSigma} by

\begin{equation}\label{eq:varianceEstimator}
     \hat \sigma_{\hat \ell_{DE}}^2(\boldsymbol{\theta}) \equiv \frac{1}{m}\sum_{k=1}^m \big(d_{u_k}(\boldsymbol{\theta})-\bar d^{(m)}(\boldsymbol{\theta})\big)^2,
\end{equation}
where $\bar d^{(m)} (\boldsymbol{\theta}) = m^{-1} \sum_{k=1}^m d_{u_k}(\boldsymbol{\theta})$, giving the estimator 
\begin{equation}\label{eq:biasCorrectDEEst}
    \hat p_{DE}(\mathbf{y} | \boldsymbol{\theta}, \mathbf{u}) \equiv \exp \Big(\hat\ell_{DE}(\mathbf{y}|\boldsymbol{\boldsymbol{\theta}})- \hat \sigma_{\hat \ell_{DE}}^2(\boldsymbol{\boldsymbol{\theta}})/2 \Big).
\end{equation}

Substituting an estimate $\hat \sigma_{\hat \ell_{DE}}^2(\boldsymbol{\boldsymbol{\theta}})$ makes the estimator in \eqref{eq:biasCorrectDEEst} only approximately unbiased, and raises the question: what do samples from a PMMH algorithm using the estimator $ \hat p_{DE}(\mathbf{y} | \boldsymbol{\boldsymbol{\theta}})$ converge to, if anything? \cite{quiroz2015speeding} note that this PMMH is still a valid MCMC on the joint ($\boldsymbol{\boldsymbol{\theta}}, \mathbf{u}$) space, but targets the density
\begin{equation}\label{eq:extended_target}
    \bar \pi (\boldsymbol{\boldsymbol{\theta}}, \mathbf{u}) = \frac{\hat p_{DE} (\mathbf{y} \vert \boldsymbol{\boldsymbol{\theta}}, \mathbf{u}) p(\mathbf{u})p(\boldsymbol{\boldsymbol{\theta}})}{\bar p(\mathbf{y})}, \text{where }
     \bar p(\mathbf{y}) = \iint_{\mathbf{u},\boldsymbol{\boldsymbol{\theta}}} \hat p_{DE} (\mathbf{y} \vert \boldsymbol{\boldsymbol{\theta}}, \mathbf{u}) p(\mathbf{u})p(\boldsymbol{\boldsymbol{\theta}}) d \mathbf{u} d \boldsymbol{\boldsymbol{\theta}}.
\end{equation}
The marginal density of $\boldsymbol{\boldsymbol{\theta}}$ is
\begin{equation}\label{eq:perturbedMarginal}
    \bar \pi (\boldsymbol{\boldsymbol{\theta}}) = \frac{\bar p (\mathbf{y} \vert \boldsymbol{\boldsymbol{\theta}})p(\boldsymbol{\boldsymbol{\theta}})}{\bar p(\mathbf{y})}, \text{where } \bar p (\mathbf{y} \vert \boldsymbol{\boldsymbol{\theta}}) \equiv \int_{\mathbf{u}} \hat p_{DE} (\mathbf{y} \vert \boldsymbol{\boldsymbol{\theta}}, \mathbf{u}) p(\mathbf{u}) d \mathbf{u}.
\end{equation}

Note that $\bar p (\mathbf{y} \vert \boldsymbol{\boldsymbol{\theta}}) \neq p (\mathbf{y} \vert \boldsymbol{\boldsymbol{\theta}})$ in general because of the (slight) bias in the likelihood estimator $\hat p_{DE} (\mathbf{y} \vert \boldsymbol{\theta}, \mathbf{u})$. This shows that PMMH based on $\hat p_{DE} (\mathbf{y} \vert \boldsymbol{\theta}, \mathbf{u})$ is still a valid MCMC scheme, but the draws $\boldsymbol{\theta}^{(1)},\ldots,\boldsymbol{\theta}^{(N)}$ target the perturbed posterior $\bar \pi (\boldsymbol{\theta})$ instead of the actual posterior $\pi (\boldsymbol{\theta})$.

Our next result from \cite{quiroz2015speeding} gives the rate at which the perturbed target $\bar \pi (\boldsymbol{\theta})$
approaches the true target posterior $\pi(\boldsymbol{\theta})$. Note that $\pi(\boldsymbol{\theta})$ depends on $n$ and $\bar \pi (\boldsymbol{\theta})$ depends on both $n$ and $m$. We make this dependence explicit by using the relevant subscripts in our asymptotic results.

\renewcommand{\labelenumi}{\roman{enumi})}
\begin{theorem}
\label{thm:perturbAsymptotics}Suppose that a PMMH algorithm is implemented
with the estimator $\hat p_{DE}(\mathbf{y} | \boldsymbol{\theta})$ in \eqref{eq:biasCorrectDEEst} using the second order parameter expanded control variates where the expansion point $\boldsymbol{\theta}^\star$ is the posterior mode, and assume that the regularity conditions in Assumption
2 in \cite{quiroz2015speeding} are satisfied. Then,
\begin{enumerate}
\item 
\[
\int_{\Theta} \left|\overline{\pi}_{m,n}(\boldsymbol{\theta})-\pi_{n}(\boldsymbol{\theta})\right| d \boldsymbol{\theta} = O\left(\frac{1}{n m^{2}}\right).
\]
\item Suppose that $h(\boldsymbol{\theta})$ is a function such that $\mathrm{E}_{\pi_{n}}[h^2(\boldsymbol{\theta})]<\infty$.
Then
\[
\left| \mathrm{E}_{\overline{\pi}_{m,n}}[h(\boldsymbol{\theta})]-\mathrm{E}_{\pi_{n}}[h(\boldsymbol{\theta})] \right| = O\left(\frac{1}{n m^{2}}\right).
\]
\end{enumerate}
\end{theorem}

Theorem \ref{thm:perturbAsymptotics} shows that the perturbation error vanishes rapidly with the subsample size at rate $O(m^{-2})$ for fixed $n$. The theorem also shows that when for example $m=O(n^{1/2})$, the perturburation error is $O(n^{-2})$. 

To analyze the scalability of the algorithm for practical work, \citet{quiroz2015speeding} make the more realistic assumption that control variates are expanded around $\boldsymbol{\theta}^\star_{\tilde n}$, the posterior mode based on a small subset of $\tilde n$ observations, rather the costly posterior mode $\boldsymbol{\theta}^\star_{n}$ based on all $n$ observations. The following corollary is proved in \citet{quiroz2015speeding}.

\begin{corollary}
\label{corr:perturbAsymptotics}Suppose that a PMMH algorithm is implemented
with the estimator $\hat p_{DE}(\mathbf{y} | \boldsymbol{\theta})$ in \eqref{eq:biasCorrectDEEst} using the second order parameter expanded control variates with expansion point $\boldsymbol{\theta}^\star_{\tilde n}$ based on a subset $\tilde n  \ll n$ of observations. Assume that $\boldsymbol{\theta}^\star_{\tilde n}-\boldsymbol{\theta}^\star_{n} = O(\tilde n^{-1/2})$, and that the regularity conditions in Assumption
2 in \cite{quiroz2015speeding} are satisfied. Then,
\begin{enumerate}
\item 
\[
\int_{\Theta} \left|\overline{\pi}_{m,n}(\boldsymbol{\theta})-\pi_{n}(\boldsymbol{\theta})\right| d \boldsymbol{\theta} = O\left(\frac{n}{m^{2}\tilde n^3}\right).
\]
\item Suppose that $h(\boldsymbol{\theta})$ is a function such that $\mathrm{E}_{\pi_{n}}[h^2(\boldsymbol{\theta})]<\infty$.
Then
\[
\left| \mathrm{E}_{\overline{\pi}_{m,n}}[h(\boldsymbol{\theta})]-\mathrm{E}_{\pi_{n}}[h(\boldsymbol{\theta})] \right| = O\left(\frac{n}{ m^{2}\tilde n^3}\right).
\]
\end{enumerate}
\end{corollary}

If $\tilde n = n^\kappa$ for some $\kappa$, then $m=O(n^{2-3\kappa})$ achieves the optimal variance of $O(1)$, and the perturbation errors in Corollary \ref{corr:perturbAsymptotics} decreases with $n$ if and only if $\kappa<2/3$. For example, if we take $\kappa = 1/2$, then $m=O(n^{1/2})$ and the posterior perturbation error is $O(n^{-1/2})$.

The asymptotics in Theorem \ref{thm:perturbAsymptotics} and Corollary \ref{corr:perturbAsymptotics} are reassuring for the method, but does not provide a practically useful way to quantify the discrepancy between $\bar \pi_{m,n}(\boldsymbol{\theta})$ and $\pi_n(\boldsymbol{\theta})$. \cite{quiroz2015speeding} derive an accurate approximation to the point-wise fractional error in the perturbed posterior distribution
\begin{equation}
    \mathrm{error}(\boldsymbol{\theta}) = \frac{\bar \pi_{m,n}(\boldsymbol{\theta})-\pi_n(\boldsymbol{\theta})}{\pi_n(\boldsymbol{\theta})}.
\end{equation}
and show that the $\mathrm{error}(\boldsymbol{\theta})$  increases with $\sigma_{\hat \ell_{DE}}^2(\boldsymbol{\theta})$ for large $\sigma_{\hat \ell_{DE}}^2(\boldsymbol{\theta})$. It is important to note however that it is only the part of $\sigma_{\hat \ell_{DE}}^2(\boldsymbol{\theta})$ that depends on $\boldsymbol{\theta}$ that affects the perturbation error; an additive constant to $\sigma_{\hat \ell_{DE}}^2(\boldsymbol{\theta})$ will give rise to a multiplicative constant to $\hat p_{DE}(\mathbf{y}|\boldsymbol{\theta})$ in \eqref{eq:biasCorrectDEEst} that also appears in $\bar p (\mathbf{y})$ and will therefore cancel in \eqref{eq:perturbedMarginal}. Hence, a large $\sigma_{\hat \ell_{DE}}^2(\boldsymbol{\theta})$ only implies a large perturbation error if $\sigma_{\hat \ell_{DE}}^2(\boldsymbol{\theta})$ varies with $\boldsymbol{\theta}$. This can be an advantage for data-expanded control variates since their errors are by construction relatively insensitive to $\boldsymbol{\theta}$, as demonstrated in Figure \ref{fig:dataexpandedAccuracy}. 

The next subsection presents an alternative approach which produces an unbiased estimator of the likelihood. Although exact, this method has two drawbacks compared to the approximate method presented in this subsection. First, the relative computational time of the algorithm is higher than the approximate method above, see Figure S8 in the supplementary material of \cite{quiroz2016exact}. Second, the exact approach can only estimate expectations of functions of the parameters, rather than the whole posterior distribution.

\subsection{Signed PMMH with the Block-Poisson estimator}\label{sec:ExactSubsampling}
The approach in the previous subsection used an unbiased estimator of the log-likelihood, which was subsequently approximately bias-corrected to estimate the likelihood 
\begin{equation} \label{eq:prodLikelihood}
    p(\mathbf{y}|\boldsymbol{\theta}) = \prod_{i=1}^n p(y_i | \boldsymbol{\theta}).
\end{equation}
We now review how to estimate this product unbiasedly using the Block-Poisson estimator proposed in \cite{quiroz2016exact}. 

The \emph{Block-Poisson} estimator is defined as 
\begin{equation}
\label{eq:PoissonEstimator}
\hat p_{\mathrm{B}}(\mathbf{y} | \boldsymbol{\theta}) = Q(\boldsymbol{\theta})\prod_{l=1}^{\lambda} \xi_l, \,\hspace{0.5cm} \xi_l = \exp\left(\frac{a+\lambda}{\lambda}\right)\prod_{h=1}^{\mathcal{X}_l}\left(\frac{\hat{d}_{m_b}^{\,\,(h,l)}-a}{\lambda}\right),
\end{equation}
where $Q(\boldsymbol{\theta})=\exp\left(\sum_{i=1}^n q_i(\boldsymbol{\theta})\right)$, with $q_i(\boldsymbol{\theta})$ being the control variates in \eqref{eq:diffEstimator}. The Block-Poisson estimator is essentially a product of $\lambda \in \mathbb{N}^+$ Poisson estimators, $\xi_l,l=1,\ldots,\lambda$ \citep{wagner1988unbiased,papaspiliopoulos2009methodological}. Each Poisson estimator in the product is based on a random number $\mathcal{X}_l \overset{indep.}{\sim}\mathrm{Pois}(1)$ of unbiased estimates $\hat {d}_{m_b}^{\,\,(h,l)}$ of $d=\sum_{i=1}^n d_i(\boldsymbol{\theta})$, i.e. the second term in the difference estimator \eqref{eq:diffEstimator}, but from a mini-batch of $m_b<m$ observations. The scalar $a\in \mathbb{R}$ is a lower bound of the $\hat {d}_{m_b}^{\,\,(h,l)}$ to ensure that $\hat p_{\mathrm{B}}(\mathbf{y} | \boldsymbol{\theta})>0$ for all $\boldsymbol{\theta}$.

\cite{quiroz2016exact} show that the Block-Poisson estimator is unbiased for the likelihood $p(\mathbf{y} | \boldsymbol{\theta})$ for all $\boldsymbol{\theta}$. The product construction in the estimator is not used for variance reduction, but to induce dependency in the subsamples over the MCMC iterations, see Section \ref{sec:CPM} below; in fact, \cite{quiroz2016exact} prove that the variance of $\hat p_{\mathrm{B}}(\mathbf{y} | \boldsymbol{\theta})$ is finite and exactly the same as the variance of the usual Poisson estimator in \citet{papaspiliopoulos2009methodological}. 

To ensure that $\hat p_{\mathrm{B}}(\mathbf{y} | \boldsymbol{\theta})$ in \eqref{eq:PoissonEstimator} is positive with probability $1$, which is necessary for PMMH, $a$ needs to be a lower bound of $\hat{d}_{m_b}$. Obtaining a lower bound is problematic for two reasons. First, a lower bound requires evaluating $d_i(\boldsymbol{\theta})$ for all data points. Second, $-a$ can be prohibitively large as the most extreme outcome of $\mathbf{u}$ needs to be covered. This is problematic because \cite{quiroz2016exact} show that $\mathrm{Var}(\hat p_{\mathrm{B}}(\mathbf{y} | \boldsymbol{\theta}))$ is minimized for $a=d-\lambda$ for any given $\lambda$. Hence, $\lambda$ must typically be very large in order for $a$ to be a lower bound, and a large $\lambda$ means many mini-batches and a high computational cost.

\cite{quiroz2016exact} instead advocate the use of a \textit{soft lower bound}, which is a lower bound resulting in $\tau \equiv \Pr(\hat p_\mathrm{B}(\mathbf{y} | \boldsymbol{\theta}) \geq 0) $ less than one, but close to it. Since the estimator might not be positive, the target cannot be defined as in \eqref{eq:extended_target}. However, further augmenting the density $\bar \pi (\boldsymbol{\theta}, \mathbf{u},s)$ with the variable $s=\mathrm{sign}(\hat p_\mathrm{B}(\mathbf{y}|\boldsymbol{\theta})) \in \{-1, 1\}$, we obtain (cf. Section \ref{sub:approxApproach})
\begin{equation}
\bar {\pi}(\boldsymbol{\theta}, \mathbf{u}, s) \equiv \frac{|\hat p_\mathrm{ B}(\mathbf{y}|\boldsymbol{\theta})|p(\boldsymbol{\theta})p(\mathbf{u})}{\tilde{p}(\mathbf{y})} = s \bar {\pi}(\boldsymbol{\theta},\mathbf{u}) ,\text{ with } \bar{\pi}(\boldsymbol{\theta},\mathbf{u}) \equiv \frac{\hat p_\mathrm{B}(\mathbf{y}|\boldsymbol{\theta})p(\boldsymbol{\theta})p(\mathbf{u})}{\tilde{p}(\mathbf{y})}, \label{eq:AugmentedPosterior}
\end{equation}
where $\tilde{p}(\mathbf{y})=\iiint_{\boldsymbol{\theta}, s, \mathbf{u}}  s \hat p_\mathrm{B}(\mathbf{y}|\boldsymbol{\theta})p(\mathbf{u})p(\boldsymbol{\theta}) d\mathbf{u} ds d\boldsymbol{\theta}$ is a normalization constant. Note that if $\tau = \Pr(s = 1) = 1$,  $\iint_{s, \mathbf{u}} \bar{\pi}(\boldsymbol{\theta},\mathbf{u}, s) d\mathbf{u} ds = \pi(\boldsymbol{\theta})$ and hence samples from the true posterior are obtained, instead of an approximation as in Section \ref{sub:approxApproach}. We argued above that $\tau=1$ is too expensive and therefore \cite{quiroz2016exact} follow \cite{lyne2015russian}, who cleverly note that
\begin{eqnarray}
\mathbb{E}_{\pi}\left(\psi\right) = \frac{\int_{\boldsymbol{\theta}} \psi(\boldsymbol{\theta})\pi(\boldsymbol{\theta})d\boldsymbol{\theta}}{\int_{\boldsymbol{\theta}} \pi(\boldsymbol{\theta}) d\boldsymbol{\theta}} 
 =  \frac{\iint_{\boldsymbol{\theta}, \mathbf{u}} \psi(\boldsymbol{\theta}) s |\hat p_\mathrm{B}(\mathbf{y}|\boldsymbol{\theta})| p(\mathbf{u})p(\boldsymbol{\theta})d\mathbf{u} d\boldsymbol{\theta}}{\iint_{\boldsymbol{\theta}, \mathbf{u}} s |\hat p_\mathrm{B}(\mathbf{y}|\boldsymbol{\theta})| p(\mathbf{u})p(\boldsymbol{\theta})d\mathbf{u} d\boldsymbol{\theta}} = \frac{\mathbb{E}_{\bar\pi}(\psi s)}{\mathbb{E}_{\bar\pi}(s)}. \label{eq:EstimationOfFunctional}
\end{eqnarray}
We can therefore obtain $N$ samples from  $\bar{\pi}(\boldsymbol{\theta}, u, s)$ in \eqref{eq:AugmentedPosterior} and estimate \eqref{eq:EstimationOfFunctional} by
\begin{eqnarray}
\widehat{\mathbb{E}}_{\pi}\left(\psi\right) & = & \frac{\sum_{i=1}^{N}\psi(\boldsymbol{\theta}^{(i)})s^{(i)}}{\sum_{i=1}^{N}s^{(i)}}, \label{eq:ISestimator}
\end{eqnarray}
that satisfies $\widehat{\mathbb{E}}_{\pi}\left(\psi\right) \overset{a.s.}{\rightarrow} {\mathbb{E}}_{\pi}\left(\psi\right)$ as $N\rightarrow\infty$. The approach in \cite{lyne2015russian} of running PMMH on the absolute posterior followed by a sign-correction by importance sampling to consistently estimate expectations of functionals is termed \emph{Signed PMMH} by \cite{quiroz2016exact}.

Under the optimal variance condition $a = d - \lambda$ it remains to choose values for the tuning parameters $\lambda$ and $m_b$. The natural approach is to choose $\lambda$ and $m_b$ to minimize a computational time similar to \eqref{eq:IACT}. \cite{quiroz2016exact} show that the computational time of Signed PMMH with the Block-Poisson estimator is
\begin{equation}\label{eq:ComputationalTimePrNonNegative}
\mathrm{CT}(\lambda, m_{b}) = m_{b}\lambda \frac{\mathrm{IACT}(\sigma^2_{\log |\hat p_{\mathrm{B}}|}(\lambda,m_b))}{(2\tau(\lambda,m_b) - 1)^2},
\end{equation}
where $\sigma^2_{\log |\hat p_{\mathrm{B}}|}(\lambda,m_b)$ is the variance of the log of the absolute value of the Block-Poisson estimator. To minimize $\mathrm{CT}(\lambda, m_{b})$ we need to compute i) IACT, ii) $\sigma^2_{\log |\hat p_{\mathrm{B}}|}(\lambda,m_b)$ and iii) $\tau(\lambda,m_b)$. All three quantities are derived in closed form in \cite{quiroz2016exact} where practical strategies for optimally tuning of $\lambda$ and $m_{b}$ to minimize CT are also proposed. The derivations are made under idealized assumptions, but the tuning is demonstrated to be near optimal. Furthermore, the guidelines for selecting $\lambda$ and $m_{b}$ are shown to be conservative in the sense of not giving too low values for $\lambda$ and $m_{b}$, which is known to be crucial in pseudo-marginal methods.

We end this subsection with a discussion of the possibility of using the Block-Poisson estimator in survey sampling, outside of a Subsampling MCMC context. We are not aware of survey sampling applications where the interest is in estimating a population product. However, the Poisson estimator is a special case of so called \textit{debiasing} estimators \citep{rhee2013unbiased}. Such estimators are useful for unbiased estimation of a quantity (e.g. the likelihood) which is a non-linear function of a quantity that can easily be estimated unbiasedly (the log-likelihood). The debiasing approach resolves this issue for general functions. It is for example possible to apply this idea to debias calibration estimators \citep{deville1992calibration} such as the ratio estimator in survey sampling.

\subsection{Dependent subsampling}\label{sec:CPM}
We have argued that controlling the variance of the log of the likelihood estimator is crucial for the efficiency of PMMH. A closer inspection of Algorithm \ref{alg:pseudoMargMH} shows that it is more correct to say it is the variance of the \emph{difference} in the log likelihood estimates at $\boldsymbol{\theta}^\prime$ and $\boldsymbol{\theta}^{(i-1)}$ that matters for PMMH. Using independent proposals for $\mathbf{u}$ makes it easy to get a gross over-estimate of the likelihood at some iteration and get stuck, as illustrated in Figure \ref{fig:subsamplingMCMC}. Refreshing only parts of the subsample in each iteration reduces the variance of the difference in the log of the estimates of the likelihood between the proposed and current point. This is achieved by making $\mathbf{u}^{(i-1)}$ (last accepted draw) and $\mathbf{u}^{\prime}$ (proposed draw) dependent. We now present two approaches from the Subsampling MCMC literature for generating dependence in $\mathbf{u}$ over the MCMC iterations, which were developed independently of the literature on repeated survey sampling for estimating changing populations over time \citep{steel2009design} in the survey sampling field. Much of this literature is focused on problems unrelated to Subsampling MCMC, for example how to avoid responders fatigue in repeated surveys, but this is certainly an area where the knowledge of survey statisticians can advance Subsampling MCMC.
\subsection*{The correlated pseudo-marginal}
\cite{deligiannidis2015correlated} present a general Correlated Pseudo-Marginal (CPM) approach to dependent particles in PMMH. Their focus is on random effects models and particle filters in state-space models where the $\mathbf{u}$ are usually Gaussian random numbers used to generate the importance samples or the particles. The correlation of the $\mathbf{u}$ over iterations is achieved by an autoregressive proposal 
\begin{equation}
\mathbf{u}^{\prime} = \phi \mathbf{u}^{(i-1)} + \sqrt{1-\phi^2}\boldsymbol{\varepsilon},
\end{equation}
where $\boldsymbol{\varepsilon} \sim N(0,\mathbf{I})$. The tuning parameter $\phi$ is set close to one to generate high persistence in the iterates of the estimated likelihood, which makes it possible to run PMMH with a variance of the log of the estimated likelihood which is roughly two orders of magnitude larger than the variance around unity in the case with independently proposed $\mathbf{u}$ \citep{deligiannidis2015correlated}.

\cite{quiroz2015speeding} apply the approach in \cite{deligiannidis2015correlated} to a subsampling context. In subsampling without replacement, $\mathbf{u}$ are binary variables with $u_i=1$ if the $i$th observation is in the subsample; see Section \ref{sec:costlyLikelihood}. \cite{quiroz2015speeding} propose using a two-state Markov Chain to generate binary dependent proposals where the transition probabilities are set to obtain the desired degree of persistence and a pre-determined expected subsample size $m^\star$. They show that this can be formulated using the same autoregressive proposal with Gaussian random variables as in \cite{deligiannidis2015correlated} using a Gaussian Copula \citep{joe2014dependence}.

\subsection*{Block pseudo-marginal}\label{sec:BPM}
\cite{quiroz2016exact} propose an alternative way to generate dependence in PMMH. For the specific problem of Subsampling MCMC without replacement, their Block Pseudo-Marginal (BPM) algorithms starts by partitioning the subsample indices $\mathbf{u} = (u_1,\ldots,u_n)$ into $G$ blocks: $\mathbf{u} = (\mathbf{u}^{(1)},\ldots,\mathbf{u}^{(G)})$. For the Block-Poisson estimator, each block consist of $\mathbf{u}$'s in one or several of the $\lambda$ products. Rather than updating all of $\mathbf{u}$ as in regular PMMH, BPM updates only one of the blocks $\mathbf{u}^{(g)}$, $g \in \{ 1,\ldots,G\}$ in each iteration, jointly with the model parameters $\boldsymbol{\theta}$. Updating only a single block in each iteration and leaving the other $G-1$ blocks unchanged makes the log-likelihood estimates highly correlated over the iterations, again making it possible to use estimators with much larger variances and still not get stuck in the PMMH. BPM is a less general approach than CPM, but has a number of advantages over CPM when it is applicable. For example, the correlation $\varrho$ between log-likelihood estimates over the iterations is, under simplifying assumptions, $1-1/G$ \citep{quiroz2016exact} and is therefore directly controlled by the number of blocks; in CPM the correlation between the logs of the estimated likelihoods is only indirectly and nonlinearly controlled by $\phi$. For subsampling, BPM offers some advantages, for example that only the $u$'s in the current block need to be generated.

\subsection{Subsampling in Hamiltonian Monte Carlo}
In Section \ref{sub:MH} we presented the Random Walk Metropolis (RWM) algorithm which proposes $\boldsymbol{\theta}$ using a random walk over the parameter space. RWM is a robust algorithm, but the local nature of RWM makes it very slow to traverse the posterior, especially in high-dimensional parameter spaces. This section presents Hamiltonian Monte Carlo, which can make much more distant proposals, and its recent extension to subsampling.

\subsection*{Hamiltonian Monte Carlo}
Hamiltonian Monte Carlo (HMC), introduced in \cite{duane1987hybrid}, is a very popular algorithm for sampling from high-dimensional posteriors; see \cite{neal2011hmc} and \cite{betancourt2017conceptual} for very accessible introductions to HMC. HMC augments the posterior $\pi(\boldsymbol{\theta})$ with fictitious momentum variables $\mathbf{m}\in \mathbb{R}^d$, of the same dimension as $\boldsymbol{\theta}$, and carries out the sampling on an extended target distribution $\bar \pi(\boldsymbol{\theta}, \mathbf{m})$. This is similar to the augmentation with the subsample indicators $\mathbf{u}$ in PMMH, but the $\mathbf{m}$ are not introduced to reduce computational cost, but to increase sampling efficiency. The momentum variables allow the algorithm to produce distant proposals while maintaining a high acceptance probability. HMC targets

\begin{equation}
    \bar\pi(\boldsymbol{\theta}, \mathbf{m}) \propto \exp(-\mathcal{H}(\boldsymbol{\theta}, \mathbf{m})),
\end{equation}
where $\mathcal{H}$ is the so called Hamiltonian, or total energy, which is here assumed to be separable in the potential ($\mathcal{U}$) and kinetic energies ($\mathcal{K}$):
\begin{equation}
    \mathcal{H}(\boldsymbol{\theta}, \mathbf{m}) = \mathcal{U}(\boldsymbol{\theta}) + \mathcal{K}(\mathbf{m}),
\end{equation}
where
\begin{equation}
     \mathcal{U}(\boldsymbol{\theta}) = -\log[ p(\mathbf{y} | \boldsymbol{\theta}) p(\boldsymbol{\theta})] \text{ and } \mathcal{K}(\mathbf{m}) = \frac{1}{2}\mathbf{m}^T \mathbf{M}^{-1} \mathbf{m},
\end{equation}
and $\mathbf{M}$ is a $d \times d$ positive definite matrix.

The HMC algorithm uses an initial momentum from $\mathbf{m}\sim N(0,\mathbf{M})$ to propagate both $\boldsymbol{\theta}$ and $\mathbf{m}$ over time $t$ along a trajectory mapped out by the Hamiltonian dynamics
\begin{align}
    \nabla_{t}\boldsymbol{\theta} &= \nabla_{\mathbf{m}} \mathcal{H}(\boldsymbol{\theta},\mathbf{m}) =  \mathbf{M}^{-1} \mathbf{m}\\
    \nabla_{t}\mathbf{m} &= -\nabla_{\boldsymbol{\theta}} \mathcal{H}(\boldsymbol{\theta},\mathbf{m}) = -\nabla_{\boldsymbol{\theta}} \mathcal{U}(\boldsymbol{\theta}), 
\end{align}
where $\nabla_t$ denotes the time derivative, and $\nabla_{\mathbf{m}}$ and $\nabla_{\boldsymbol{\theta}}$ are the gradients with respect to $\mathbf{m}$ and $\boldsymbol{\theta}$, respectively.
Hamiltonian dynamics has several very attractive properties \citep{neal2011hmc}, one of them being that it keeps the Hamiltonian conserved: $\nabla_{t}\mathcal{H} = 0$. Hamiltonian dynamics can therefore be used to generate proposals for $\boldsymbol{\theta}$ over long distances that are accepted with probability one. In practical computer implementations, however, one needs to discretize the Hamiltonian dynamics, so the total energy is not preserved and we need a MH accept/reject step with acceptance probability less than one. The most common way to discretize the Hamiltonian dynamics in HMC is the leapfrog method \citep{neal2011hmc}. Algorithm \ref{alg:HMC} outlines the complete HMC algorithm using the leapfrog method. 

The performance of HMC is very sensitive to its two tuning parameters, the leapfrog step size $\epsilon$ and number of leapfrog steps $L$. The No-U-Turn algorithm proposed by \cite{hoffman2014no} is an effective method to tune $\epsilon$ and $L$.

\subsection*{Hamiltonian Monte Carlo with Energy Conserving Subsampling}

HMC is a very efficient algorithm that scales well to high-dimensional posteriors, but it needs to repeatedly evaluate the gradient $\nabla_{\boldsymbol{\theta}} \mathcal{U}(\boldsymbol{\theta})$ at each of the $L$ leapfrog iterations in every MH iteration. Note that $L$ typically needs to be rather large if we want to make distant moves without too much energy loss ($\epsilon$ small). Usually $\nabla_{\boldsymbol{\theta}} \mathcal{U}(\boldsymbol{\theta})$ is costly whenever $\mathcal{U}(\boldsymbol{\theta})$ is costly, so the same computational hurdles discussed for standard MH apply also to HMC, but now to a much larger extent because of the $L$ gradient evaluations in the leapfrog steps, see Algorithm \ref{alg:HMC} in Appendix \ref{sec:algorithms}. Several authors have proposed running the leapfrog iterations on a subsample of the data to speed up computations \citep{neal2011hmc,chen2014stochastic,betancourt2015fundamental}, thereby replacing $\nabla_{\boldsymbol{\theta}} \mathcal{U}(\boldsymbol{\theta})$ by an unbiased subsample estimate. However, such an approach strips HMC of its energy conserving property and distant proposals tend to be rejected with high probability \citep{betancourt2015fundamental}. The energy loss comes from using a subsample estimate of the Hamiltonian dynamics that no longer operates on the true Hamiltonian used in the accept/reject step.

\cite{dang2017hamiltonian} observe that this disconnect between the dynamics and the Hamiltonian can be easily avoided by extending the Subsampling MCMC algorithm in \cite{quiroz2015speeding} to HMC proposals. The Energy Conserving Subsampling (HMC-ECS) algorithm in \cite{dang2017hamiltonian} samples from the extended target 
\begin{equation}
    \bar \pi (\boldsymbol{\theta}, \mathbf{m}, \mathbf{u}) \propto \exp \big( -\hat{\mathcal{H}}(\boldsymbol{\theta}, \mathbf{m}, \mathbf{u}) \big) p(\mathbf{u} ),
\end{equation}
where $p(\mathbf{u})$ is the distribution for the subsample selection indicators. The Hamiltonian in HMC-ECS is based on a subsample estimate of the Hamiltonian

\begin{equation}
    \hat{\mathcal{H}}(\boldsymbol{\theta}, \mathbf{m}, \mathbf{u}) = \hat{\mathcal{U}}(\boldsymbol{\theta}, \mathbf{u}) + \mathcal{K}(\mathbf{m}),
\end{equation}
where 
\begin{equation}\label{eq:biasedPotentialEstimator}
   \hat{\mathcal{U}}(\boldsymbol{\theta}, \mathbf{u}) = -\Big( \hat \ell_{DE}(\mathbf{y}|\boldsymbol{\theta} , \mathbf{u}) - \frac{1}{2} \hat \sigma_{\hat \ell_{DE}}^2 + \log p(\boldsymbol{\theta}) \Big) \text{ and } \mathcal{K}(\mathbf{m}) = \frac{1}{2}\mathbf{m}^T \mathbf{M}^{-1} \mathbf{m}.
\end{equation}
The potential energy estimator in \eqref{eq:biasedPotentialEstimator} makes HMC-ECS target a (slightly) perturbed posterior distribution, whose error can be controlled by the theory in \cite{quiroz2015speeding}.

Algorithm \ref{alg:HMC-ECS} gives the HMC-ECS algorithm. This sampler is a so called two-block Metropolis-within-Gibbs sampler which iteratively samples from the two full conditional posterior distributions
\begin{itemize}
    \item $\mathbf{u} \sim \bar \pi(\mathbf{u} | \boldsymbol{\theta}, \mathbf{m})$ using Metropolis-Hastings
    \item $(\boldsymbol{\theta}, \mathbf{m}) \sim \bar \pi(\boldsymbol{\theta}, \mathbf{m} | \mathbf{u} )$ using HMC.
\end{itemize}

Following the usual HMC algorithm, the gradient used for generating proposal trajectories in the update for $(\boldsymbol{\theta}, \mathbf{m})$ in HMC-ECS is with respect to the target $\mathcal{\hat U}(\boldsymbol{\theta},\mathbf{u})$. 

The key feature of HMC-ECS is using the \textit{same} subsample $\mathbf{u}$ to estimate the Hamiltonian and to generate the trajectories in the Hamiltonian dynamics. Thus, HMC-ECS conserves the energy exactly as in the original HMC. As an example, \cite{dang2017hamiltonian} compares HMC and HMC-ECS in a big data application on firm bankruptcy in a logistic additive spline model with a $d=89$-dimensional posterior. \cite{dang2017hamiltonian} report that HMC-ECS gives an effective sample size that is up to three orders of magnitude larger than HMC for a given time budget. Moreover, the average acceptance probability of HMC-ECS is $79.3\%$, which is only marginally lower than the $81.8\%$ for HMC.

\section{Concluding remarks}
We have presented the pseudo-marginal approach to subsampling in Markov Chain Monte Carlo from the perspective of a survey statistician. We have reviewed several effective control variates for variance reduction of the likelihood estimator which make Subsampling MCMC scalable to large datasets. We have also presented methods for correlating the subsamples over the MCMC iterations, ultimately leading to algorithms that allow much more variable likelihood estimators. Much of the focus was given to unbiased estimators of the log-likelihood and methods for bias-correction of the resulting likelihood estimators. Focusing on the log-likelihood gives a direct analogy to estimating the population total, a long studied problem in survey sampling. This comes at the cost of giving an algorithm that samples from a slightly perturbed posterior, and we also review an alternative approach with unbiased likelihood estimators that can be used to obtain exact posterior expectations of functions of the parameters. We hope that this review makes it easier for survey statisticians to enter the field of Subsampling MCMC, and that it inspires them to make contributions to further enhance the efficiency of the algorithms.

\section{Acknowledgements}
Matias Quiroz and Robert Kohn were partially supported by Australian Research Council Center of Excellence grant CE140100049.

\bibliographystyle{apalike}
\bibliography{sankhyaRef}

\newpage

\appendix

\section{Algorithms}\label{sec:algorithms}
This appendix contains the main sampling algorithms discussed in the paper.

\begin{algorithm}
\SetAlgoLined
\KwInput{data \textbf{y}, likelihood function $p(\mathbf{y} | \boldsymbol{\theta})$, prior density $p(\boldsymbol{\theta})$, proposal density $q(\boldsymbol{\theta}^\prime \vert \boldsymbol{\theta})$, random number generator for $q(\boldsymbol{\theta}^\prime \vert \boldsymbol{\theta})$, initial value $\boldsymbol{\theta}^{(0)}$, number of iterations $N$.} 
\BlankLine
\For{$i = 1$ \KwTo $N$}{
   draw $\boldsymbol{\theta}^\prime \sim q(\cdot \vert \boldsymbol{\theta}^{(i-1)})$ \\
   set $\boldsymbol{\theta}^{(i)} \leftarrow \boldsymbol{\theta}^\prime$ with probability \\
   \hspace{2cm}$\alpha = \min \Big( 1, \frac{p(\mathbf{y} | \boldsymbol{\theta}^\prime)p(\boldsymbol{\theta}^\prime)}{p(\mathbf{y} | \boldsymbol{\theta}^{(i-1)})p(\boldsymbol{\theta}^{(i-1)})}  \frac{q(\boldsymbol{\theta}^{(i-1)} \vert \boldsymbol{\theta}^\prime)}{q(\boldsymbol{\theta}^\prime \vert \boldsymbol{\theta}^{(i-1)})} \Big)$ \\
   else set $\boldsymbol{\theta}^{(i)} \leftarrow \boldsymbol{\theta}^{(i-1)}$
} 
\BlankLine
\KwOutput{autocorrelated random draws $\boldsymbol{\theta}^{(1)},\ldots,\boldsymbol{\theta}^{(N)}$ from $\pi(\boldsymbol{\theta}) \propto p(\mathbf{y} | \boldsymbol{\theta})p(\boldsymbol{\theta})$.}
\BlankLine
\caption{The Metropolis-Hastings algorithm\label{alg:MH}}
\end{algorithm}

\begin{algorithm}
\SetAlgoLined
\KwInput{data \textbf{y}, unbiased likelihood estimator $ \hat p(\mathbf{y} | \boldsymbol{\theta}, \mathbf{u})$, prior density $p(\boldsymbol{\theta})$, proposal density $q(\boldsymbol{\theta}^\prime \vert \boldsymbol{\theta})$, random number generator for $q(\boldsymbol{\theta}^\prime \vert \boldsymbol{\theta})$, initial value $\boldsymbol{\theta}^{(0)}, \mathbf{u}^{(0)}$, random number generator for the augmentation variables $\mathbf{u}$, number of augmentation variables $m$, number of iterations $N$.} 
\BlankLine
\For{$i = 1$ \KwTo $N$}{
   generate $\mathbf{u}^\prime \sim p(\mathbf{u})$ \\
   generate $\boldsymbol{\theta}^\prime \sim q(\cdot \vert \boldsymbol{\theta}^{(i-1)})$ \\
   set $(\boldsymbol{\theta}^{(i)},\mathbf{u}^{(i)}) \leftarrow (\boldsymbol{\theta}^\prime,\mathbf{u}^\prime)$ with probability \\
   \hspace{2cm}$\alpha = \min \Big( 1, \frac{\hat p(\mathbf{y} | \boldsymbol{\theta}^{\prime}, \mathbf{u}^\prime)p(\boldsymbol{\theta}^\prime)}{\hat p(\mathbf{y} | \boldsymbol{\theta}^{(i-1)}, \mathbf{u}^{(i-1)})p(\boldsymbol{\theta}^{(i-1)})}  \frac{q(\boldsymbol{\theta}^{(i-1)} \vert \boldsymbol{\theta}^\prime)}{q(\boldsymbol{\theta}^\prime \vert \boldsymbol{\theta}^{(i-1)})} \Big)$ \\
   else set $(\boldsymbol{\theta}^{(i)},\mathbf{u}^{(i)}) \leftarrow (\boldsymbol{\theta}^{(i-1)},\mathbf{u}^{(i-1)})$
} 
\BlankLine
\KwOutput{autocorrelated random draws $\boldsymbol{\theta}^{(1)},\ldots,\boldsymbol{\theta}^{(N)}$ from $\pi(\boldsymbol{\theta}) \propto p(\mathbf{y} | \boldsymbol{\theta})p(\boldsymbol{\theta})$.}
\BlankLine
\caption{The pseudo-marginal Metropolis-Hastings algorithm\label{alg:pseudoMargMH}}
\end{algorithm}

\begin{algorithm}
\SetAlgoLined
\KwInput{data \textbf{y}, $p(\mathbf{y} | \boldsymbol{\theta}, \mathbf{u})$, prior density $p(\boldsymbol{\theta})$, initial value $\boldsymbol{\theta}^{(0)}$, step size $\epsilon$, number of leapfrog steps $L$, number of iterations $N$.} 
\BlankLine
define $\mathcal{U}(\boldsymbol{\theta}) = -\log [p(\mathbf{y} | \boldsymbol{\theta})p(\boldsymbol{\theta})]$ \\
define $\mathcal{K}(\mathbf{m}) = \frac{1}{2}\mathbf{m}^T \mathbf{M}^{-1} \mathbf{m}$ \\
\For{$i = 1$ \KwTo $N$}{
  
   \BlankLine
   \textbackslash\textbackslash\hspace{0.1cm} generate trajectory by $L$ leapfrog steps \\
   draw initial momentum $\mathbf{m}\sim N(0,\mathbf{M})$ \\
   set $\boldsymbol{\theta}^{\prime} \leftarrow \boldsymbol{\theta}^{(i-1)}$ \\
   set $\mathbf{m}^{\prime} \leftarrow \mathbf{m} - \frac{\epsilon}{2} \nabla_{\boldsymbol{\theta}} \mathcal{U}(\boldsymbol{\theta}^{\prime})$ \\
   \For{$l = 1$ \KwTo $L$}{
        set $\boldsymbol{\theta}^{\prime} \leftarrow \boldsymbol{\theta}^{\prime} + \epsilon \mathbf{M}^{-1} \mathbf{m}^{\prime}$ \\
        \uIf{$l \neq L$}{$\mathbf{m}^{\prime} \leftarrow \mathbf{m}^{\prime} -\epsilon {\nabla_{\boldsymbol{\theta}} \mathcal{U}(\boldsymbol{\theta}^{\prime})}$}
        \Else{$\mathbf{m}^{\prime} \leftarrow \mathbf{m}^{\prime} -\frac{\epsilon}{2} {\nabla_{\boldsymbol{\theta}} \mathcal{U}(\boldsymbol{\theta}^{\prime})}$}
    } 
    \BlankLine
    
    \textbackslash\textbackslash\hspace{0.1cm} accept or reject $(\boldsymbol{\theta}^{\prime},\mathbf{m}^{\prime})$ \\
    set $\boldsymbol{\theta}^{(i)} \leftarrow \boldsymbol{\theta}^\prime$ with probability \\
   \hspace{2cm}$\alpha = \min \Big[ 1, \exp \big( { -\mathcal{U}(\boldsymbol{\theta}^{\prime})}+\mathcal{U}(\boldsymbol{\theta}^{(i-1)}) 
   -\mathcal{K(\mathbf{m}^{\prime})}+\mathcal{K(\mathbf{m})} \big) \Big] $ \\
   else set $\boldsymbol{\theta}^{(i)} \leftarrow \boldsymbol{\theta}^{(i-1)}$
   \BlankLine
} 
\BlankLine
\KwOutput{autocorrelated random draws $\boldsymbol{\theta}^{(1)},\ldots,\boldsymbol{\theta}^{(N)}$ from $\pi(\boldsymbol{\theta}) \propto p(\mathbf{y} | \boldsymbol{\theta})p(\boldsymbol{\theta})$.}
\BlankLine
\caption{Hamiltonian Monte Carlo (HMC)}\label{alg:HMC}
\end{algorithm}

\begin{algorithm}
\SetAlgoLined
\KwInput{data \textbf{y}, unbiased likelihood estimator $\hat p(\mathbf{y} | \boldsymbol{\theta}, \mathbf{u})$, prior density $p(\boldsymbol{\theta})$, initial value $\boldsymbol{\theta}^{(0)}$, initial subsample $\mathbf{u}^{(0)}$, random number generator for $\mathbf{u}$, step size $\epsilon$, number of leapfrog steps $L$, number of iterations $N$.} 
\BlankLine
define $\mathcal{U}(\boldsymbol{\theta},\mathbf{u}) = -\log [p(\mathbf{y} | \boldsymbol{\theta}, \mathbf{u})p(\boldsymbol{\theta})]$ \\
define $\mathcal{K}(\mathbf{m}) = \frac{1}{2}\mathbf{m}^T \mathbf{M}^{-1} \mathbf{m}$ \\
\For{$i = 1$ \KwTo $N$}{
    \BlankLine
    \textbackslash\textbackslash\hspace{0.1cm} update the subsample $\mathbf{u}$ \\
    generate $\mathbf{u}^{\prime} \sim p(\mathbf{u})$ \\
    set $\mathbf{u}^{(i)} \leftarrow \mathbf{u}^{\prime}$ with probability \\
    \hspace{2cm}$\alpha_{\mathbf{u}} = \min \Big( 1, \frac{\hat p(\mathbf{y} | \boldsymbol{\theta}^{(i-1)}, \mathbf{u}^{\prime})}{\hat p(\mathbf{y} | \boldsymbol{\theta}^{(i-1)}, \mathbf{u}^{(i-1)})}   \Big)$ \\
    else set $\mathbf{u}^{(i)} \leftarrow \mathbf{u}^{(i-1)}$
  
   \BlankLine
   \textbackslash\textbackslash\hspace{0.1cm} generate trajectory by $L$ leap frog steps \\
   draw initial momentum $\mathbf{m}\sim N(0,\mathbf{M})$ \\
   set $\boldsymbol{\theta}^{\prime} \leftarrow \boldsymbol{\theta}^{(i-1)}$ \\
   set $\mathbf{m}^{\prime} \leftarrow \mathbf{m} - \frac{\epsilon}{2} \nabla_{\boldsymbol{\theta}} \mathcal{U}(\boldsymbol{\theta}^{\prime},\mathbf{u}^{(i)})$ \\
   \For{$l = 1$ \KwTo $L$}{
        set $\boldsymbol{\theta}^{\prime} \leftarrow \boldsymbol{\theta}^{\prime} + \epsilon \mathbf{M}^{-1} \mathbf{m}^{\prime}$ \\
        \uIf{$l \neq L$}{$\mathbf{m}^{\prime} \leftarrow \mathbf{m}^{\prime} -\epsilon \nabla_{\boldsymbol{\theta}} \mathcal{U}(\boldsymbol{\theta}^{\prime},\mathbf{u}^{(i)})$}
        \Else{$\mathbf{m}^{\prime} \leftarrow \mathbf{m}^{\prime} -\frac{\epsilon}{2} \nabla_{\boldsymbol{\theta}} \mathcal{U}(\boldsymbol{\theta}^{\prime},\mathbf{u}^{(i)})$}
    } 
    \BlankLine
    
    \textbackslash\textbackslash\hspace{0.1cm} accept or reject $(\boldsymbol{\theta}^{\prime},\mathbf{m}^{\prime})$ \\
    set $\boldsymbol{\theta}^{(i)} \leftarrow \boldsymbol{\theta}^\prime$ with probability \\
   \hspace{2cm}$\alpha = \min \Big[ 1, \exp \big(  -\mathcal{U}(\boldsymbol{\theta}^{\prime},\mathbf{u}^{(i)})+\mathcal{U}(\boldsymbol{\theta}^{(i-1)},\mathbf{u}^{(i)}) 
   -\mathcal{K(\mathbf{m}^{\prime})}+\mathcal{K(\mathbf{m})} \big) \Big] $ \\
   else set $\boldsymbol{\theta}^{(i)} \leftarrow \boldsymbol{\theta}^{(i-1)}$
   \BlankLine
} 
\BlankLine
\KwOutput{autocorrelated random draws $\boldsymbol{\theta}^{(1)},\ldots,\boldsymbol{\theta}^{(N)}$ from $\pi(\boldsymbol{\theta}) \propto p(\mathbf{y} | \boldsymbol{\theta})p(\boldsymbol{\theta})$.}
\BlankLine
\caption{HMC with Energy Conserving Subsampling (HMC-ECS)\label{alg:HMC-ECS}}
\end{algorithm}

\section{Details for the Poisson regression example}
This appendix gives the details for the control variates in our illustrative Poisson regression example. \cite{quiroz2015speeding} gives general expressions for the gradients and Hessians in the GLM class, and provides general compact expression that reduces the computational complexity of the control variates.

\subsection*{The Poisson regression model}
The Poisson regression is of the form
\begin{equation*}
    y_i | \mathbf{x}_i, \boldsymbol{\theta} \overset{indep.}{\sim} \mathrm{Pois}(\lambda_i), \hspace{0.5cm} \lambda_i = \exp(\alpha + \mathbf{x}_i^T \beta),
\end{equation*}
where $\boldsymbol{\theta} = (\alpha,\beta)^T$.

\subsection*{Parameter-expanded control variates}

Let $\mathbf{w}_i = (1,\mathbf{x}_i^T)^T$. The log-likelihood contribution from the $i$th observation is 
\begin{equation*}
    \ell_i(\boldsymbol{\theta}) = y_i \mathbf{w}_i^T \boldsymbol{\theta} - \exp(\mathbf{w}_i^T \boldsymbol{\theta}) - \log(y_i!) 
\end{equation*}   
with gradient and Hessian
\begin{equation*}
    \nabla_{\boldsymbol{\theta}} \ell_i(\boldsymbol{\theta}) = (y_i  - \exp(\mathbf{w}_i^T \boldsymbol{\theta}))\mathbf{w}_i  
\end{equation*}   
\begin{equation*}
    \nabla_{\boldsymbol{\theta} \boldsymbol{\theta}^T}^2 \ell_i(\boldsymbol{\theta}) = - \exp(\mathbf{w}_i^T \boldsymbol{\theta})\mathbf{w}_i\mathbf{w}_i^T
\end{equation*} 

Let $\mu(\boldsymbol{\theta},\mathbf{x}) = \alpha + \mathbf{x}^T\beta = \mathbf{w}^T \boldsymbol{\theta}$. The parameter-expanded control variate in \eqref{eq:paramControlVariates} is then 

\begin{align*} 
    \ell_i(\boldsymbol{\theta}) &\approx y_i \mu( \boldsymbol{\hat \theta},\mathbf{x}_i) - \exp(\mu( \boldsymbol{\hat \theta},\mathbf{x}_i)) - \log(y_i!)  \\
    &+  [y_i  - \exp(\mu( \boldsymbol{\hat \theta},\mathbf{x}_i))](\mu_i(\theta)-\mu_i(\boldsymbol{\hat \theta})) \\
    &-  \frac{1}{2}\exp(\mu(\boldsymbol{\hat \theta},\mathbf{x}_i))(\mu(\theta,\mathbf{x}_i)-\mu(\boldsymbol{\hat \theta},\mathbf{x}_i))^2,
\end{align*}

\subsection*{Data-expanded control variates}

The log-likelihood contribution from the $i$th observation is 
\begin{equation*}
    \ell_i(\theta) = y_i (\alpha + \mathbf{x}_i^T \beta) - \exp(\alpha + \mathbf{x}_i^T \beta) - \log(y_i!) 
\end{equation*}   
with gradient and Hessian
\begin{equation*}
    \nabla_{y_i} \ell_i(\theta) = \alpha + \mathbf{x}_i^T \beta - \psi_0(y_i+1),
\end{equation*}  
where $\psi_k(z) = \nabla_z^k \log \Gamma(z)$ is the polygamma function of order $k$,
\begin{align*}
    \nabla_{\mathbf{x}_i} \ell_i(\theta) &= (y_i - \exp(\alpha + \mathbf{x}_i^T \beta)) \beta, \hspace{0.1cm}
    \nabla_{y_i y_i}^2 \ell_i(\theta) = - \psi_1(y_i+1), \\
    \nabla_{\mathbf{x}_i \mathbf{x}_i^T}^2 \ell_i(\boldsymbol{\theta}) &= - \exp(\alpha + \mathbf{x}_i^T \beta) \beta \beta^T, \text{ and } 
    \nabla_{y_i \mathbf{x}_i^T}^2 \ell_i(\boldsymbol{\theta}) = \beta.
\end{align*} 
 
We can write the gradients and Hessian compactly by defining $\mathbf{z}_i=(y_i,\mathbf{x}_i^T)^T$,
\begin{equation*}
\nabla_{z_i} \ell_i(\boldsymbol{\theta}) =
    \begin{bmatrix}
        \alpha + \mathbf{x}_i^T \beta - \psi_0(y_i+1)       \\
        (y_i - \exp(\alpha + \mathbf{x}_i^T \beta)) \beta       
    \end{bmatrix}
\end{equation*}

\begin{equation*}
\nabla_{z_i z_i^T}^2 \ell_i(\boldsymbol{\boldsymbol{\theta}}) =
    \begin{bmatrix}
        - \psi_1(y_i+1)       & \beta^T  \\
        \beta                 & - \exp(\alpha + \mathbf{x}_i^T \beta) \beta \beta^T
    \end{bmatrix}.
\end{equation*}

Let $\mu(\boldsymbol{\theta}, \mathbf{x}) = \alpha + \mathbf{x}^T \beta$. The data-expanded control variate in \eqref{eq:dataControlVariates} can after some simplifications be expressed as

\begin{align*} 
    \ell_i(\boldsymbol{\theta}) &\approx y_{c_i} \mu(\boldsymbol{\theta}, \mathbf{x}_{c_i}) - \exp(\mu(\boldsymbol{\theta}, \mathbf{x}_{c_i})) - \log(y_{c_i}!)  \\
    &+  (y_i - y_{c_i})(\mu(\boldsymbol{\theta}, \mathbf{x}_{c_i}) - \psi_0(y_{c_i}+1)) -\frac{1}{2} (y_i - y_{c_i})^2\psi_1(y_{c_i}+1)  \\  
    &+[y_i-\exp(\mu(\boldsymbol{\theta}, \mathbf{x}_{c_i}))] (\mu(\boldsymbol{\theta}, \mathbf{x}_i)-\mu(\boldsymbol{\theta}, \mathbf{x}_{c_i})  )  \\
    &-\frac{1}{2} \exp(\mu(\boldsymbol{\theta}, \mathbf{x}_{c_i}))(\mu(\boldsymbol{\theta}, \mathbf{x}_i)-\mu(\boldsymbol{\theta}, \mathbf{x}_{c_i})  )^2  .
\end{align*}

\end{document}